\documentclass[11pt]{article} 
\usepackage{graphicx,amssymb,mathrsfs,array,multirow}  
\begin{document}  
 
\vskip 30pt

\begin{center}  
{\Large{\bf Ameliorating the popular lepton mixings with A4 symmetry: A see-saw model for realistic neutrino masses and mixing}}\\
\vspace*{1cm}  
\renewcommand{\thefootnote}{\fnsymbol{footnote}}  
{ {\sf Soumita Pramanick$^{1,2}$\footnote{email: soumita509@gmail.com}}
} \\  
\vspace{10pt}  
{\small  {\em $^1$Harish-Chandra Research Institute, Chhatnag Road, Jhunsi,
Allahabad 211019, India \\
$^2$Department of Physics, University of Calcutta,  
92 Acharya Prafulla Chandra Road, Kolkata 700009, India\\
}}
\normalsize  

\normalsize

\end{center}  

\begin{abstract} 
\textit{
A model for neutrino masses and mixing is presented using the see-saw mechanism.
The model combines Type -I and Type-II see-saw contributions
of which the latter dominates. 
The scalars and the leptons in the model are assigned 
$A4$ charges suitable to obtain the mass matrices required for the scheme.
The Type -II see-saw accommodates atmospheric mass splitting 
and maximal mixing in the atmospheric sector ($\theta_{23}=\pi/4$). It is characterized by vanishing 
solar mass splitting 
and $\theta_{13}$ whereas the third neutrino mixing angle can acquire any value, $\theta_{12}^0$.
Particular alternatives of $\theta_{12}^0$ viz. $\theta_{12}^0=35.3^\circ$
(tribimaximal), $45.0^\circ$  (bimaximal),
$31.7^\circ$ (golden ratio) are accounted for. Another choice of $\theta_{12}^0=0^\circ$ (no solar mixing)
is also considered. Incorporating the corrections provided by the subdominant Type-I see-saw involves degenerate perturbation theory 
due to vanishing solar splitting in the Type -II see-saw
enabling the solar mixing angle to receive substantial corrections. Apart from amending the solar
sector the Type-I see-saw also
tunes all the neutrino oscillation
parameters into the allowed ranges thus interrelating them all. 
Thus the model is testable in the light of future experimental data. As an example, $\theta_{23}$ emerges in the first (second) octant 
for normal (inverted) ordering. 
CP-violation is controlled by phases present in
the right-handed Majorana neutrino mass matrix, $M_{\nu R}$.
Only normal ordering is allowed if these phases are absent. 
If $M_{\nu R}$ is complex the Dirac CP-violating phase $\delta$,
can be large, i.e., $\sim \pm \pi/2$, and inverted ordering is also allowed. T2K 
and NOVA preliminary data favouring 
normal ordering and $\delta \sim -\pi/2$ predicts lightest neutrino mass to be 0.05 eV or
more within the model framework.
}

\vskip 5pt \noindent  
\texttt{PACS No:~ 14.60.Pq}  \\  
\texttt{Key Words:~~Neutrino mixing, $\theta_{13}$, Solar
splitting, A4, see-saw, Leptonic CP-violation}
\end{abstract}  

\renewcommand{\thesection}{\Roman{section}} 
\setcounter{footnote}{0} 
\renewcommand{\thefootnote}{\arabic{footnote}} 
\noindent

\section{Introduction}
Intensive experimental investigations worldwide
have determined neutrino masses and mixing to a great extent. In spite of these 
neutrinos retain certain mysteries including
the ordering of their masses, their absolute mass scale, their Dirac or Majorana nature, the octant of the atmospheric 
mixing angle $\theta_{23}$ and CP-violation in lepton sector. While future experiments
address these riddles, here a model of neutrino masses and mixing
in concord with the experimental observations is proposed.
The two small quantities $\theta_{13}$ and the
ratio, $R \equiv \Delta m^2_{solar}/\Delta m^2_{atmos}$ can get interrelated
when both are derived from a single perturbation \cite{br}.
In \cite{pr} larger mixing parameters like $\Delta
m^2_{atmos}$ and $\theta_{23} = \pi/4$ were ascribed to the dominant
fundamental structure of neutrino masses and mixing whereas the other oscillation parameters
 i.e., $\theta_{13}, \theta_{12}$, the deviation
of $\theta_{23}$ from $\pi/4$, and $\Delta m^2_{solar}$
originated from a smaller see-saw \cite{seesaw} generated perturbation\footnote{Earlier attempts on neutrino mass models with
some oscillation parameters much smaller than the others can be located in
\cite{old}.}. This induces constraints on the measured parameters.
Certain symmetries can give rise to vanishing $\theta_{13}$ rather easily
and new models based on perturbations of such structures are also common in literature
\cite{models, LuhnKing}.
\vskip 1pt
Here, a schematic outline of the current exercise is given.
The following standard parametrization of the lepton mixing matrix -- the
Pontecorvo, Maki, Nakagawa, Sakata (PMNS) matrix -- $U$ has been used
\begin{eqnarray}
U = \left(
          \begin{array}{ccc}
          c_{12}c_{13} & s_{12}c_{13} & s_{13}e^{-i\delta}  \\
- c_{23}s_{12} + s_{23}s_{13}c_{12}e^{i\delta} & c_{23}c_{12} +
s_{23}s_{13}s_{12}e^{i\delta}&  s_{23}c_{13}\\
 s_{23}s_{12} + c_{23}s_{13}c_{12}e^{i\delta}&  -s_{23}c_{12} +
c_{23}s_{13}s_{12}e^{i\delta} & c_{23}c_{13} \end{array} \right)
\;\;,
\label{PMNS}
\end{eqnarray}
where $c_{ij} = \cos \theta_{ij}$ and $s_{ij} = \sin
\theta_{ij}$.
Neutrino masses and mixing are generated by a 
two-component Lagrangian, one of the 
dominant Type-II see-saw kind while the subdominant contribution 
originates from Type-I see-saw. The
larger atmospheric mass splitting, $\Delta m^2_{atmos}$ and maximal atomspheric
mixing ($\theta_{23}=\pi/4$) is embedded within the Type-II see-saw structure
whereas the solar splitting, $\Delta m^2_{solar}$ and $\theta_{13}$ are kept to be
zero. The solar mixing angle can vary continuously
and acquire any desired value of $\theta_{12}^0$.
Needless to mention that neither $\Delta m^2_{solar}$ nor $\theta_{13}$ are vanishing \cite{t13_s3}.
Evidences of non-maximal yet large $\theta_{23}$ exist. The solar mixing angle $\theta_{12}$ is also constrained by experiments.
The Type-I see-saw alleviates all these issues. Since the solar splitting is vanishing in the Type-II 
see-saw scenario, the first two mass eigenstates are degenerate.
In order to lift this degeneracy with the help
of Type-I see-saw contribution one has to use
degenerate perturbation theory. As a consequence of this, corrections to the 
solar mixing angle can be large.
\vskip 1pt
The starting structure can be of tribimaximal (TBM), bimaximal (BM), and golden
ratio (GR) mixings. All of these have $\theta_{13} = 0$  
and $\theta_{23} = \pi/4$, $\theta_{12}^0$ being the only discriminating factor 
as specified in Table \ref{t1}. In this Table, the fourth option
corresponds to no solar mixing
(NSM) i.e., $\theta_{12}^0=0$ which has the virtue of the
mixing angles to be either maximal, i.e.,  $\pi/4$ ($\theta_{23}$) or vanishing
($\theta_{13}$ and $\theta_{12}^0$). 
An $A4$-based model with identical objectives only for the NSM case 
was studied in \cite{ourA4}.
This attempt along with \cite{ourA4} differ from the other earlier works on $A4$
\cite{A4mr, A4af, otherA4} as in most of them neutrino mass matrix was derived
as an outcome of a Type-II see-saw mechanism and obtaining TBM was
of chief importance. Recent activities directed towards more realistic
mixing patterns \cite{newMa} often leading to 
breaking of $A4$ symmetry  can be found in \cite{tanimoto}.
\vskip 1pt
A few distinctive aspects of this model are worth noting at this point. 
Firstly, a combination of Type-I and Type-II see-saw is considered.
Secondly, the model is constructed to accommodate many popular mixing 
patterns. 
This is the first attempt of this kind using $A4$ flavour symmetry that
amends several popular lepton mixing patterns in a single stroke
in which Type-II see-saw is the dominant contribution whereas Type-I see-saw
is the subdominant component.
The symmetries are broken 
spontaneously. Further, soft symmetry breaking terms are prohibited. All 
symmetry conserving terms are included in the Lagrangian. 
Scalars and leptons involved in the model are assigned
suitable $A4$ charges to implement this feature.
An analogous pursuit based on $S3\times Z3$ resulted in \cite{ourS3}.

\begin{table}[tb]
\begin{center}
\begin{tabular}{|c|c|c|c|c|}
\hline
  Model &TBM &BM & GR & NSM\\ \hline
$\theta^0_{12}$ & 35.3$^\circ$ & 45.0$^\circ$  & 31.7$^\circ$ &
0.0$^\circ$ \\ \hline
\end{tabular}
\end{center}
\caption{\sf{$\theta^0_{12}$ for different popular lepton mixing 
patterns viz. TBM, BM, and GR  mixing. NSM represents the case with
vanishing solar mixing. }}
\label{t1}
\end{table}
\vskip 1pt
All the three neutrino mixing angles and the solar mass splitting
receives first order corrections from a single source -- the Type -I see-saw in this model.
Owing to the common origin, they all get interrelated. These correlations
are characteristic features of this particular model. Indeed the model has 
a large number of parameters, but it must noted that only the region of the
parameter space allowed by the neutrino mass and mixing data obeying these correlations
is considered.
\vskip 1pt
An analysis of the model initiates the discussion. In the next section, the operational
strategy is described. The results so obtained are compared to the experimental data
in the following section, succeeded by the conclusions and inferences of this work.
Some essential ideas of the of the discrete symmetry $A4$ are presented in Appendix \ref{GroupA4}.
A detailed study of the rich scalar sector to the extent of local minimization of the scalar potential is furnished in Appendix \ref{V_A4}. In Appendix \ref{MassA4} algebraic details of the mass matrix calculations while going to the flavour basis of the neutrinos from the Lagrangian basis can be found.

\section{The Mass Model}
The model comprises of scalars and leptons
with specific $A4$ charges. 
All terms allowed by the symmetries under consideration are included in the Lagrangian.
No soft symmetry-breaking term is included.
 
\begin{table}[t]
\begin{center}
\begin{tabular}{|c|c|c|c|c|}
\hline
Fields & Notations & $A4$ & $SU(2)_L$ ($Y$) & $L$   \\ 
 \hline
 & & &  &   \\
Left-handed leptons&$(\nu_i,l_i)_L$&3&2 (-1)& 1 \\
 & & &   &  \\
\hline
 & & &   &  \\
 &$l_{1R} $ &1& &   \\
 & & & &    \\
\cline{2-3}
 & & & &    \\
Right-handed charged leptons & $l_{2R} $&$1'$&1 (-2) & 1
\\
 & & &  &   \\
\cline{2-3}
 & & &  &   \\
 & $l_{3R} $&$1''$& &     \\
 & & &  &   \\
\hline
Right-handed  neutrinos& $N_{iR} $&3&1 (0) & -1 \\
\hline
\end{tabular}
\end{center}
\caption{\sf The lepton catalogue of the model. The 
$A4$ quantum numbers assignments of the fields are featured together with their $SU(2)_L$ properties. 
The hypercharge, 
$Y$, and lepton number, $L$, are displayed.}
\label{tab1f}
\end{table}
The right-handed charged leptons transform
as $1 (e_R)$, $1'
(\mu_R)$, and $1'' (\tau_R)$ under $A4$.
The left-handed lepton doublets of 
three flavours constitute an $A4$ triplet, so does
the right-handed neutrinos\footnote{The notation followed closely resembles that of \cite{A4mr}.}.  
Table \ref{tab1f} shows the lepton constituents 
of the model together with their transformation properties under $A4$ and $SU(2)_L$.
The hypercharge and lepton number assignments are also shown\footnote{
Opposite lepton numbers are assigned to $\nu_L$ and $N_R$ in order to prohibit their 
coupling with $\Phi$ so that the Dirac mass matrix can remain proportional to the identity matrix.
}.
The choices of $A4$ properties of the fields are not unique. A list of all possible options can be found in \cite{rb} of which this model adopts
class B. The model is restricted to leptons only\footnote{
Quark models based on $A4$ has been explored in \cite{quarks1} and \cite{quarks2}.}.

\begin{table}[t]
{\tiny
\begin{center}
\begin{tabular}{|c|c|c|c|c|c|}
\hline
Purpose & Notations & $A4$ & $SU(2)_L$ & $L$ & $vev$ \\ 
 &  &  & ($Y$) & &  \\  \hline
 & & & & &  \\
Charged fermion mass& $\Phi=\pmatrix{\phi_1^+ & \phi_1^0\cr
\phi_2^+& \phi_2^0\cr\phi_3^+& \phi_3^0}$&3&2 (1) &0&$\langle\Phi\rangle=
\frac{v}{\sqrt{3}} \pmatrix{0 & 1 \cr 0 & 1 \cr 0 & 1 }$ \\
 & & & &  &  \\
 \hline
 & & & & &  \\
 Neutrino Dirac mass& $\eta=(\eta^0 ,\eta^- )$
&1&2 (-1)&2& $\langle\eta\rangle=\pmatrix{u,0}$
\\
 & & & & &  \\
\hline
 & & & &  & \\
Type-II see-saw mass& $\hat{\Delta}^L_a=\pmatrix{
\hat{\Delta}^{++}_{1a} & \hat{\Delta}^{+}_{1a} & \hat{\Delta}^0_{1a} \cr
 \hat{\Delta}^{++}_{2a} & \hat{\Delta}^{+}_{2a} & \hat{\Delta}^0_{2a} \cr
 \hat{\Delta}^{++}_{3a} & \hat{\Delta}^{+}_{3a} & \hat{\Delta}^0_{3a} }^L$
&3&3 (2)&-2&$\langle \hat{\Delta}^L_a \rangle= v_{La} \pmatrix{0
& 0 & 1  \cr
0 & 0 & 0  \cr
0 & 0 & 0 }$ 
\\
 & & & & &  \\
\hline
 & & & &  & \\
Type-II see-saw mass& $\hat{\Delta}^L_b=\pmatrix{
\hat{\Delta}^{++}_{1b} & \hat{\Delta}^{+}_{1b} & \hat{\Delta}^0_{1b} \cr
 \hat{\Delta}^{++}_{2b}& \hat{\Delta}^{+}_{2b} & \hat{\Delta}^0_{2b}\cr
 \hat{\Delta}^{++}_{3b} & \hat{\Delta}^{+}_{3b} & \hat{\Delta}^0_{3b} }^L$
&3&3 (2)&-2&$\langle \hat{\Delta}^L_b \rangle= v_{Lb} \pmatrix{0
& 0 & 1  \cr
0 & 0 & 1  \cr
0 & 0 & 1 }$ 
\\
 & & & & &  \\
 \hline
 & & 1 & 3 (2) &-2& $\langle\Delta_1^L\rangle=\pmatrix{0,0,u_L}$
\\
\cline{3-6}
 Type-II see-saw mass &$\Delta_\zeta^L =
(\Delta_\zeta^{++},\Delta_\zeta^+,\Delta_\zeta^0)^L$  &$1'$&3 (2) &-2&
$\langle\Delta_2^L\rangle=\pmatrix{0,0,u_L}$
\\
\cline{3-6}
 & &$1''$&3 (2) &-2& $\langle\Delta_3^L\rangle=\pmatrix{0,0,u_L}$
\\
\hline
 & & & & &  \\
 Right-handed neutrino mass& $\hat{\Delta}^R_a=\pmatrix{
 \hat{\Delta}^0_{1a}\cr
 \hat{\Delta}^0_{2a} \cr
 \hat{\Delta}^0_{3a} }^R
$&3&1 (0)&2&$\langle \hat{\Delta}^R_a \rangle= v_{Ra} \pmatrix{1 \cr
 1 \cr  1 }$ 
\\
 & & & & &   \\
\hline
 & & & & &  \\
 Right-handed neutrino mass& $\hat{\Delta}^R_b=\pmatrix{
 \hat{\Delta}^0_{1b}\cr
 \hat{\Delta}^0_{2b} \cr
 \hat{\Delta}^0_{3b} }^R
$&3&1 (0)&2&$\langle \hat{\Delta}^R_b \rangle= v_{Rb} \pmatrix{1 \cr
 \omega\cr  \omega^2  }$ 
\\
 & & & & &   \\
\hline
 & & & & &  \\
 Right-handed neutrino mass& $\hat{\Delta}^R_c=\pmatrix{
 \hat{\Delta}^0_{1c}\cr
 \hat{\Delta}^0_{2c} \cr
 \hat{\Delta}^0_{3c} }^R
$&3&1 (0)&2&$\langle \hat{\Delta}^R_c \rangle= v_{Rc} \pmatrix{1 \cr
 \omega^2 \cr  \omega }$ 
\\
 & & & & &   \\
 \hline
 & & & &  & \\
 Right-handed neutrino mass
&$\Delta_1^R = (\Delta_1^{0})^R$  &$1$&1 (0) &2&
$\langle\Delta_1^R\rangle={u}_{1R}$ \\ 
& & & & &  \\
 \hline
 & & & &  & \\
 Right-handed neutrino mass
&$\Delta_2^R = (\Delta_2^{0})^R$  &$1'$&1 (0) &2&
$\langle\Delta_2^R\rangle={u}_{2R}$ \\ 
& & & & &  \\
 \hline
 & & & &  & \\
 Right-handed neutrino mass
&$\Delta_3^R = (\Delta_3^{0})^R$  &$1''$&1 (0) &2&
$\langle\Delta_3^R\rangle={u}_{3R}$ \\ 
& & & & &  \\
\hline
\end{tabular}
\end{center}
\caption{\sf The scalar sector of the model. 
The $A4$ charges as well as the $SU(2)_L$ nature of the scalars
are exhibited. The hypercharge
,$Y$, lepton number, $L$, and the vacuum
expectation value (vev) configurations of the scalars are also presented.} 
\label{tab1s}
}
\end{table}

Masses of all leptons originate from $A4$-invariant Yukawa
couplings. Several scalar fields have to be included\footnote{
Models addressing this issue by separating the breaking of $SU(2)_L$ and
$A4$ are widely studied in literature \cite{A4af}.
The former is mediated by the 
usual doublet and triplet scalars of $SU(2)_L$ that are invariant 
under $A4$. The breaking of $A4$ is induced by the {\em vev}
of `flavon' scalar fields that are singlets of $SU(2)_L$ but their transformations 
under $A4$ is non-trivial. 
Though such models are economic effective dimension-5
interactions comes into play in order to connect the fermions with the 
two types of scalar fields simultaneously leading to an interpretation as an effective theory.} that acquire suitable 
vacuum expectation values ({\em vev}s). The strategy of choosing the scalar field multiplets requires some elaboration. An idea of the mass matrices of the left- and right-handed neutrinos in the flavour basis (charged lepton mass matrix diagonal) that are suitable for our avowed goal can be acquired from our previous work \cite{ourS3}.  The Lagrangian is written down in a basis which is unitarily related to the flavour basis. Consequently, the mass matrices in this defining basis have somewhat complicated structures for which the motivation is not initially obvious. These forms of the mass matrices (below) arise from a rather large set of scalars and their {\em vev}s.

\vskip 1pt
The charged leptons acquire their masses through the $SU(2)_L$ doublet scalar fields $\Phi_i ~(i=1,2,3)$
forming an $A4$ triplet. The neutrino Dirac mass matrix is generated by an $A4$ invariant $SU(2)_L$
doublet $\eta$, having lepton number 2. $SU(2)_L$ triplet scalars are required for the Type-II
see-saw for left-handed neutrino mass matrix that include $A4$ triplet fields $\hat{\Delta}_a^L$
and $\hat{\Delta}_b^L$ along with $\Delta_\zeta^L, ~\zeta=1,2,3$ transforming as $1$, $1'$, $1''$
of $A4$. These are used to construct the dominant Type-II see-saw neutrino mass matrix.
Effects of the subdominant Type-I see-saw contribution is included perturbatively. $A4$ conserving
Yukawa couplings produce the right-handed neutrino
mass matrix as well. Several $SU(2)_L$
singlet scalars are involved in generation of the Majorana masses for the right-handed neutrinos
viz. $\hat{\Delta}^R_p$ ($p=a,b,c$) transforming as $A4$ triplets and $\Delta^R_\gamma$ ($\gamma=1,2,3$)
transforming as $1$, $1'$ and $1''$ under $A4$. Table \ref{tab1s} evinces transformation properties
of the model scalars under $A4$ and $SU(2)_L$ together with their hypercharge, lepton number and 
{\em vev} configurations.
The {\em vev}s of the $SU(2)_L$ doublet scalars are of ${\cal O}(M_W)$ while that of the $SU(2)_L$ triplets are several orders of  
magnitude smaller than the doublet {\em vev}s in concord 
with the small neutrino masses as well as the $\rho$ 
parameter of electroweak symmetry breaking. As expected, the {\em vev}s of the $SU(2)_L$ singlets
responsible for right-handed neutrino mass lies much above the electroweak scale.
The mass terms of the neutrinos (both Type-I and Type-II see-saw) and that of the 
charged leptons are generated by a $SU(2)_L\times U(1)_Y$ conserving
Lagrangian that preserves $A4$ as well\footnote{Lepton number is also conserved for the mass terms of Dirac kind.}:
\begin{eqnarray}
\mathscr{L}_{mass}&=& y_j \rho_{jik} \bar{l}_{Li} l_{Rj}\Phi_{k}^{0} 
 ~~{\rm(charged ~lepton ~mass)}  \nonumber\\
&+& f \rho_{1ik} \bar{\nu}_{Li}  N_{Rk} \eta^0 
 ~~{\rm(neutrino ~Dirac ~mass)} \nonumber\\
&+&\frac{1}{2}\left(\sum_{n=a,b} \hat{Y}^L_n ~\alpha_{ijk}\nu_{Li}^TC^{-1}\nu_{Lj}
\hat{\Delta}_{nk}^{L0}
+Y^L_{\zeta} ~\rho_{\zeta ij}\nu_{Li}^TC^{-1}\nu_{Lj}\Delta_\zeta^{L0}\right)
 ~~{\rm(neutrino ~Type\!-\!II ~see\!-\!saw ~mass)} \nonumber \\
&+&  \frac{1}{2}\left(\sum_{p=a,b,c}\hat{Y}^R_p ~\alpha_{ijk} N_{Ri}^TC^{-1}N_{Rj}
\hat{\Delta}_{kp}^{R0}
+Y^R_{\gamma} ~\rho_{\gamma ij} N_{Ri}^TC^{-1} N_{Rj}\Delta_\gamma^{R0}\right) 
~~{\rm(rh ~neutrino ~mass)} + h.c.  
\label{e1}
\end{eqnarray}
The scalars acquire the following {\em
vev}s ($SU(2)_L$ part is suppressed):
\begin{equation}
\langle \Phi^0 \rangle = \frac{v}{\sqrt{3}} \pmatrix{1 \cr 1 \cr 1} \;,\;
\langle \eta^0 \rangle = u  \;,
\langle \hat{\Delta}^{L0}_a \rangle = v_{La}\pmatrix{1 \cr 0 \cr 0} \;,\; 
\langle \hat{\Delta}^{L0}_b \rangle = v_{Lb}\pmatrix{1 \cr 1 \cr 1} \;,\; 
\langle \Delta_1^{L0} \rangle =  
\langle \Delta_2^{L0} \rangle =  
\langle \Delta_3^{L0} \rangle = {u}_L \;, 
\label{vev1}
\end{equation}

\begin{equation} 
\langle \hat{\Delta}^{R0}_a \rangle = 
v_{Ra}\pmatrix{1 \cr 1 \cr 1} \;,\; 
\langle \hat{\Delta}^{R0}_b \rangle = 
v_{Rb}\pmatrix{1 \cr \omega \cr \omega^2} \;,\;
\langle \hat{\Delta}^{R0}_c \rangle = 
v_{Rc}\pmatrix{1 \cr \omega^2 \cr \omega} \;,\;
\label{vev2}
\end{equation}

\begin{equation}
\langle \Delta_1^{R0} \rangle = {u}_{1R} \;,\; 
\langle \Delta_2^{R0} \rangle = {u}_{2R} \;,\; 
\langle \Delta_3^{R0} \rangle = {u}_{3R} \;.\; 
\label{vev3}
\end{equation}
An elaborate study of the $A4$ conserving scalar potential involving the fields 
listed in Table \ref{tab1s} is presented in Appendix \ref{V_A4} of this paper. 
Local minimization is performed and the conditions corresponding to the 
particular {\em vev} structures as indicated in Eqs. (\ref{vev1}-\ref{vev3})
are obtained.
\vskip 1pt
The mass matrix for the charged
leptons and the left-handed Majorana neutrinos so obtained are:
\begin{equation}
M_{e\mu\tau}=\frac{v}{\sqrt{3}}
\pmatrix{y_1 & y_2 & y_3\cr
y_1 & \omega y_2 & \omega^2 y_3\cr
y_1 & \omega^2 y_2 & \omega y_3} \;,\;M_{\nu L}=
\pmatrix{(Y^L_1+ 2 Y^L_2)u_L & \frac{1}{2}\hat{Y}^L_b v_{Lb}& \frac{1}{2}\hat{Y}^L_b v_{Lb} \cr
\frac{1}{2}\hat{Y}^L_b v_{Lb} & (Y^L_1 - Y^L_2)u_L & \frac{1}{2}(\hat{Y}^L_a v_{La} +\hat{Y}^L_b v_{Lb} )\cr
\frac{1}{2}\hat{Y}^L_b v_{Lb} & \frac{1}{2}(\hat{Y}^L_a v_{La} +\hat{Y}^L_b v_{Lb} ) & (Y^L_1 - Y^L_2)u_L},
\label{mmatrix1}
\end{equation}
where the choice of $Y^L_2 = Y^L_3$ is made. The Yukawa couplings involved in the 
charged lepton mass matrix satisfies $y_1 v = m_e, ~ y_2 v = m_\mu , ~ y_3 v =
m_\tau$.
The neutrino mass matrix of Dirac nature and the 
right-handed neutrino mass matrix of Majorana kind acquires the following structures:
\begin{equation}
M_D = f u ~\mathbb{I}\;\;,\;\; 
M_{\nu R}= m_R\pmatrix{\chi_1& 
\chi_6 &\chi_5\cr 
\chi_6& \chi_2 & \chi_4\cr 
\chi_5 &\chi_4 & \chi_3}\;\;.
\label{mmatrix2}
\end{equation}
$m_D$ sets the scale of Dirac masses of the neutrinos
where one can identify $fu=m_D$.
The scale of the Type-II see-saw neutrino masses is much smaller than that of the charged leptons
i.e., ${\cal O}(M_{\nu L})\sim u_L,v_{La},v_{Lb} $ where $u_L,v_{La},v_{Lb} << v$.  Such a possibility 
that the triplet {\em vev} is much smaller than the doublet {\em vev} can be obtained as shown in \cite{MaAndSarkar}, albeit in a model with fewer scalars.
The scale of the right-handed Majorana neutrino masses 
is set by $m_R$ 
and $\chi_i$ in  Eq. (\ref{mmatrix2}) are dimensionless quantities\footnote{See Appendix \ref{MassA4} for exact expressions of $\chi_i$ in  Eq. (\ref{mmatrix2}).} of ${\cal O}(1)$. 
\vskip 1pt
The mass matrices in Eq. (\ref{mmatrix1}) could be expressed in 
a more convenient form by applying a couple of transformations.
The non-hermitian charged lepton mass matrix can be diagonalised by applying a transformation $U_L$ (below) on  the left-handed lepton doublets and no transformation on the right-handed charged leptons. 
The transformation matrices 
are expressed as:
\begin{equation}
U_L = {1 \over{\sqrt 3}}
\pmatrix{1 & 1 & 1 \cr
1 & \omega^2 &  \omega\cr
1 &  \omega&  \omega^2}\;.
\label{chngbasis}
\end{equation}
This basis in which the charged lepton mass matrix is diagonal and the entire 
lepton mixing is governed by the neutrino sector is termed as the {\em flavour basis} in which the
mass matrices acquire the following forms:
\begin{equation}
M_{e\mu\tau}^{flavour} = \pmatrix{m_e & 0 & 0 \cr 0 & m_\mu & 0 \cr
0 & 0 & m_\tau}\;\;,\;\;  
M_{\nu L}^{flavour} = {1 \over 2}
\pmatrix{2m^{(0)}_1 & 0 & 0 \cr 0 & m^+ & m^- \cr 0 & m^- & m^+} \;\;.
\label{mflav1}
\end{equation}
Here $m^\pm\equiv m_3^{(0)}\pm m_1^{(0)}$.
Therefore, $m^-$ is positive (negative) for normal (inverted) ordering.
As noted earlier, $M_{\nu L}^{flavour}$, which arises from 
the Type-II see-saw, is the dominant contribution to the neutrino mass.
\vskip 1pt
Demanding that the neutrino Dirac mass matrix, which couples the left- and right-handed neutrinos,  preserves its proportionality to the identity
matrix necessitates that the transformation applied on the right-handed neutrino fields must be $V_R=U_L$.  Thus we get,
\begin{equation}
M_D = f u ~\mathbb{I}\;\;,\;\;  
M_{\nu R}^{flavour} =(V_R^\dagger
\,M_{\nu R}\,V_R^\dagger)={m_R \over 4ab}
\pmatrix{r_{11} & r_{12} & r_{13} \cr r_{12}& r_{22} & r_{23} \cr r_{13} & r_{23} & r_{33}} 
\;\;.
\label{mflav1a}
\end{equation} 
The matrices in Eq. (\ref{mflav1a}) will take part in the
Type-I see-saw mechanism\footnote{Explicit forms of $r_{ij}$ in Eq. (\ref{mflav1a})
can be found in Appendix \ref{MassA4}.}. Various identification of the products of the Yukawa couplings and the {\em vev}s
with the neutrino mass and mixing parameters are necessary for the mass
matrices to be expressed in the forms as presented in Eqs. (\ref{mflav1}) and (\ref{mflav1a}). Appendix
\ref{MassA4} comprises of these algebraic details.

\section{Modus Operandi}
The four mass matrices in the {\em flavour basis} obtained from the model are given in Eq. (\ref{mflav1})
and (\ref{mflav1a}). In this basis the entire lepton mixing and CP-violation is
controlled solely by the neutrino sector to which we restrict our discussion now onwards.
The Type-II see-saw
derived $M_{\nu L}^{flavour}$ is the dominant component to which the subdominant contribution attributed
by the Type-I see-saw is incorporated by perturbation theory.
The {\em flavour basis} mass matrices have to undergo one more basis transformations for successful implementation 
of this scheme. More precisely they ought to be expressed in the {\em mass basis} of the neutrinos
which by definition has the left-handed neutrino mass matrix diagonal in it. Thus,
\begin{equation}
M^0 = M_{\nu L}^{mass} = U^{0T} M_{\nu L}^{flavour} U^0 = 
\pmatrix{m^{(0)}_1 & 0 & 0 \cr
0 & m^{(0)}_1 & 0 \cr 0 & 0 & m^{(0)}_3} \;\;,
\label{mass0}
\end{equation}
where,
\begin{equation}
U^0=
\pmatrix{\cos \theta_{12}^0 & \sin \theta_{12}^0  & 0 \cr -\frac{\sin
\theta_{12}^0}{\sqrt{2}} & \frac{\cos \theta_{12}^0}{\sqrt{2}} &
{1\over\sqrt{2}} \cr
\frac{\sin \theta_{12}^0}{\sqrt{2}} & -\frac{\cos
\theta_{12}^0}{\sqrt{2}}  & {1\over\sqrt{2}}}.
\label{mix0}
\end{equation}
The left-handed neutrino fields in the mass basis ($|\nu_L^{mass}\rangle$) are connected to the ones in the
flavour basis ($|\nu_L^{flavour}\rangle$) by this $U^0$ furnished in Eq. (\ref{mix0}). One can obtain the
$|\nu_L^{mass}\rangle$ by applying $U^{0\dagger}$ on $|\nu_L^{flavour}\rangle$
i.e., $|\nu_L^{mass}\rangle=U^{0\dagger}|\nu_L^{flavour}\rangle$.
It immediately follows from Eqs. (\ref{mass0}), (\ref{PMNS}) and (\ref{mix0}) that in the
Type-II see-saw component solar splitting is absent, $\theta_{13} = 0$
 and $\theta_{23} = \pi/4$. The columns of $U^0$ are the unperturbed flavour basis.
\vskip 1pt
Once again we demand that in the mass basis the neutrino Dirac mass matrix remains proportional to identity.
In order to satisfy this the same transformation ($U^{0\dagger}$) has to be applied on the 
right-handed neutrino fields. This leads to changes in form of right-handed neutrino
mass matrix given by $M_{\nu R}^{mass}=(U^{0\dagger}\, M_{\nu R}^{flavour}\,U^0)$. The matrices contributing in Type-I see-saw are as follows:
\begin{equation}
M_D=m_D\mathbb{I}\ \ {\rm and} \ \ M_{\nu R}^{mass} =\frac{m_R}{2\sqrt{2}ab}
\pmatrix{0 & b& b \cr b & \frac{a}{\sqrt{2}}&-\frac{a}{\sqrt{2}} 
\cr b & -\frac{a}{\sqrt{2}}
& \frac{a}{\sqrt{2}}}  \  \ .
\label{mflavgen}
\end{equation}
Here $a$ and $b$ are dimensionless quantities\footnote{See Eq. (\ref{rij}) in Appendix \ref{MassA4} for details.} of ${\cal O}(1)$.
It is imperative to note that $a$ and $b$ can in general be complex.
One can in principle trade off $a$ and $b$ in terms of complex numbers  
$y e^{-i\phi_2}$ and $x e^{-i\phi_1}$ respectively,
where $x$ and $y$ are dimensionless real quantities of ${\cal O}(1)$.
The Type-I see-saw contribution so obtained is given by:
\begin{equation}
 M' =  \left[M_D^T(M_{\nu R})^{-1}M_D \right]  
= {m_D^2 \over m_R} 
\pmatrix{0 & y~e^{i\phi_1} & y~e^{i\phi_1} \cr 
y~e^{i\phi_1} & {x~e^{i\phi_2} \over  \sqrt 2} 
& {-x~e^{i\phi_2} \over \sqrt 2} \cr 
y~e^{i\phi_1} & {-x~e^{i\phi_2} \over \sqrt 2} & 
{x~e^{i\phi_2} \over \sqrt 2}}\;\;.
\label{pertmat}
\end{equation}
Here the Dirac mass matrix is proportional to identity.
It was checked that the same results can follow as long as
$M_D$ is diagonal. $M_{\nu R}^{mass}$
exhibits a $N_{2R}\leftrightarrow N_{3R}$ discrete symmetry. The results
remain intact even if that choice is relaxed. Now onwards the entire
procedure is carried on in the {\em mass basis} of the neutrinos
using the mass matrices expressed in Eqs. (\ref{mass0}) and (\ref{pertmat}).
\vskip 1pt
The method followed below essentially consists of the following steps. Form the Type-II see-saw a lepton mixing of the form of Eq. (\ref{mix0}) is generated, with $\theta_{12}^0$ of any preferred value. At this stage, only the atmospheric mass splitting is non-zero and atmospheric mixing is maximal. Next, the Type-I see-saw is included using degenerate perturbation theory. The solar mass splitting and the desired $\theta_{12}$ are first obtained. Then the third column of the mixing matrix is calculated and compared 
with Eq. (\ref{PMNS}) to extract $\theta_{13}$, $\theta_{23}$, and $\delta$.

\section{Results}
The neutrino mass matrices derived from Type-I and Type-II see-saw mechanism have been 
discussed in the previous section, of which the former is 
significantly smaller than the latter. In absence of the Type-I see-saw
contribution the leptonic
mixing matrix is characterized by $\theta_{13} = 0$, $\theta_{23} =
\pi/4$, and $\theta_{12}^0$ is free to vary. Consequences for 
four choices of the value of $\theta_{12}^0$ corresponding 
to TBM, BM, GR, and NSM cases together with vanishing solar splitting are examined.
This along with the atmospheric mass splitting allowed by the data depict
the Type-II see-saw structure. Inclusion of Type-I see-saw corrections
perturbatively up to first order modulates the neutrino oscillation
parameters into the ranges preferred by data. 
Owing to the vanishing solar splitting in the Type-II see-saw contribution
the first two mass eigenstates are degenerate.
Thus in the solar sector degenerate perturbation theory
has to be applied. Hence the first order corrections to the solar mixing angle
can be large.
The global best-fit of the oscillation
parameters are displayed in the next section.

\subsection{Data}
\label{sec:data}
The current 3$\sigma$ global fits of the neutrino oscillation parameters are:
\cite{Gonzalez_s3, Valle_s3} 
\begin{eqnarray}
\Delta m_{21}^2 &=& (7.02 - 8.08) \times 10^{-5} \, {\rm eV}^2, \;\;
\theta_{12} = (31.52 - 36.18)^\circ, \nonumber \\
|\Delta m_{31}^2| &=& (2.351 -  2.618) \times 10^{-3}
\, {\rm eV}^2, \;\;
\theta_{23} = (38.6 - 53.1)^\circ \,, \nonumber \\
\theta_{13} &=& (7.86 - 9.11)^\circ, \;\; \delta =
(0 - 360)^\circ \;\;.
\label{results}
\end{eqnarray}
These numbers are taken from NuFIT2.1 of 2016 \cite{Gonzalez_s3}. Needless to mention, 
$\Delta m_{ij}^2 \equiv m_i^2 - m_j^2$, such that $\Delta m_{31}^2 > 0$
for normal ordering (NO) and $\Delta m_{31}^2 < 0$ for inverted 
ordering (IO). Two
best-fit points of $\theta_{23}$ are evinced by the data in the first and in the second
octants. Towards the end of the paper it is discussed how the model
can accommodate the recent 
T2K and NOVA hints \cite{T2K_s3, Nova_s3} of $\delta$ close 
to -$\pi/2$.

\subsection{Real $M_{\nu R}$ ($\phi_1 = 0
~{\rm or} ~\pi, \phi_2 = 0 ~{\rm or} ~\pi$)}\label{sec3}
As a warm-up exercise let us consider the simpler case of
$M_{\nu R}$ real. In such a scenario there is no CP-violation
as the phases $\phi_{1,2}$ of Eq. (\ref{pertmat}) are 0 or $\pi$.
This leads to four different alternatives available for choosing
$\phi_{1}$ and $\phi_{2}$. These are captured compactly by taking
$x$ and $y$ real and allowing them to assume both signs for 
notational convenience. It will be soon clear how the experimental
observations prefer one or the other of these four alternatives.
Thus for real $M_{\nu R}$ the Type -I see-saw contribution appears like:
\begin{equation}
 M' = 
{m_D^2 \over  m_R} 
\pmatrix{0 & y & y \cr y & {x \over  \sqrt 2} & -{x \over \sqrt 2} \cr 
y & -{x \over \sqrt 2} & {x \over \sqrt 2}}\;\;.
\label{pert1}
\end{equation}
The degeneracy of the two neutrino masses in the Type-II see-saw
ensuring the vanishing solar splitting necessitates the application
of degenerate perturbation theory to obtain the corrections 
for the solar sector mixing parameters \footnote
{Since degenerate perturbation theory is
used in the solar sector, the first order correction to the solar
mixing angle $\zeta$ is not constrained to be small.}.
The entire dynamics of this sector is dictated by the upper  
$2\times2$ submatrix of $M'$ given by:
\begin{equation}
M'_{2\times2} = {m_D^2 \over  m_R} 
\pmatrix{0 & y \cr y & {x/\sqrt 2}} \;.
\label{solr}
\end{equation}
This gives rise to:
\begin{equation}
\theta_{12} = \theta_{12}^0 + \zeta \;\;,\;\; \tan 2\zeta= 2 \sqrt 2
\left(\frac{y}{x}\right)  \; .
\label{solangr}
\end{equation}
\begin{table}[tb]
\begin{center}
\begin{tabular}{|c|c|c|c|c|}
\hline
Model ($\theta^0_{12}$) &TBM (35.3$^\circ$) &BM (45.0$^\circ$) &
GR (31.7$^\circ$)  & NSM (0.0$^\circ$) \\ \hline
$\zeta$  & -4.0$^\circ \leftrightarrow$ 0.6$^\circ$ & -13.7$^\circ
\leftrightarrow$  -9.1$^\circ$  & -0.4$^\circ \leftrightarrow$
4.2$^\circ$ & 31.3$^\circ \leftrightarrow$  35.9$^\circ$ \\ \hline 
$\epsilon$  & -4.0$^\circ \leftrightarrow$  0.6$^\circ$ &
-14.5$^\circ \leftrightarrow$  -9.3$^\circ$  & -0.4$^\circ
\leftrightarrow$ 4.2$^\circ$ & 44.0$^\circ \leftrightarrow$
56.7$^\circ$ \\ \hline 
$\epsilon -  \theta_{12}^0$ & -39.2$^\circ
\leftrightarrow$  -34.6$^\circ$ & -59.5$^\circ
\leftrightarrow$  -54.4$^\circ$  & -39.2$^\circ \leftrightarrow$
-30.0$^\circ$ & 44.0$^\circ \leftrightarrow$  56.7$^\circ$ \\
\hline
\end{tabular}
\end{center}
\caption{\sf{Data allowed $3\sigma$ ranges of $\zeta$
(Eq. (\ref{solangr})), $\epsilon$ (Eq. (\ref{etar})), and $(\epsilon -
\theta_{12}^0)$ for different popular
mixing patterns are shown. }}
\label{tlim}
\end{table}

For functional ease it is useful to define a quantity, $\epsilon$ as:
\begin{equation}
\sin\epsilon = \frac{y}{\sqrt{y^2 + x^2/2}}\;\;{\rm and} ~\cos\epsilon
= \frac{x/\sqrt{2}}{\sqrt{y^2 + x^2/2}}\;\;,\;\;{\rm i.e.,} ~\tan \epsilon =
\frac{1}{2} \tan 2\zeta \;.
\label{etar}
\end{equation} 
Once a mixing pattern is selected, the corresponding
$\theta_{12}^0$ gets fixed and the experimental
bounds of $\theta_{12}$  determines the $3\sigma$ ranges of $\zeta$
and $\epsilon$ by means of Eq. (\ref{results}) and
Eq. (\ref{etar}) as featured in Table \ref{tlim}. 
The ratio $(y/x)$ is positive (negative) when $\zeta$ is positive (negative).
From Eq. (\ref{etar}) it is evident that the 
sign of $y$ is regulated by the value of $\epsilon$. 
Putting all these facts together it is easy to infer
that $x$ is positive always,
or in other words $\phi_2$ must be 0, while $y$ has to be
positive, $\phi_1 = 0$ (negative, $\phi_1 = \pi$) for NSM (BM).
In case of TBM and GR, both signs of $y$
are admissible.
The solar splitting provided by the Type-I see-saw as extracted 
from Eq. (\ref{solr}) is: 
\begin{equation}
\Delta m^2_{solar}=  {\sqrt{2} m_D^2 \over m_R} ~m^{(0)}_1
\sqrt{x^2 + 8y^2} =  {\sqrt{2} m_D^2 \over m_R} ~m^{(0)}_1
\frac{x}{\cos 2\zeta} \;\;. 
\label{solspltr}
\end{equation}

For the mass basis form of the mass matrix in Eq. (\ref{mass0}), 
the mixing in the leptonic sector is completely
given by the $U^0$ given in Eq. (\ref{mix0}). After including
the Type-I see-saw correction to the mass matrices there is a further
contribution to the mixing matrix as well, now given by:
\begin{equation}
U = U^0 U_\nu \; {\rm where} ~U_\nu=
\pmatrix{\cos \zeta & -\sin \zeta  & \kappa_r \sin\epsilon \cr 
\sin\zeta & \cos \zeta & -\kappa_r \cos\epsilon \cr
\kappa_r \sin (\zeta - \epsilon) & \kappa_r \cos (\zeta - \epsilon) & 1} \;,
\label{mixr}
\end{equation}
with
\begin{equation}
\kappa_r \equiv {m_D^2 \over {m_R m^-}} \sqrt{y^2 + x^2/2} = 
{m_D^2 \over {m_R m^-}} \frac{x}{\sqrt{2} \cos\epsilon}\;\;.
\label{kappa}
\end{equation}

The third column of the lepton mixing matrix is:
\begin{equation}
|\psi_3\rangle =
\pmatrix{\kappa_r \sin (\epsilon - \theta_{12}^0) \cr 
{1\over \sqrt 2}{[1 - \kappa_r \cos (\epsilon - \theta_{12}^0)]} 
\cr {1\over \sqrt 2}{[1 + \kappa_r \cos (\epsilon - \theta_{12}^0)]} } \ \ \ .
\label{psi3_1}
\end{equation} 
As already pointed out, $x$ is always positive, $\kappa_r$ is positive
(negative)  for NO (IO). 

Eq. (\ref{psi3_1}) when mapped to
the third column of Eq. (\ref{PMNS}) leads to:
\begin{equation}
\sin \theta_{13}\cos\delta = \kappa_r \sin (\epsilon -
\theta_{12}^0) \;\;, 
\label{s13r}
\end{equation}
and 
\begin{equation}
\tan (\pi/4 - \theta_{23}) \equiv \tan \omega = \kappa_r \cos
(\epsilon - \theta_{12}^0) \;.
\label{omegar}
\end{equation}
The allowed ranges of $(\epsilon - \theta_{12}^0)$
for the different mixing patterns is given in Table
\ref{tlim} . The CP-phase $\delta$ is 0 ($\pi$) 
when $\sin (\epsilon - \theta_{12}^0)$ is
positive (negative)
in case of normal ordering\footnote{Inverted ordering is 
prohibited 
for real $M_{\nu R}$.}. It can be immediately concluded that $\delta = 0$ for the NSM
from Table \ref{tlim} and $\delta = \pi$ for the rest of the options under
study. CP is conserved for both the values of $\delta$.

\begin{figure}[tbh]
\begin{center}
\includegraphics[scale=0.75,angle=0]{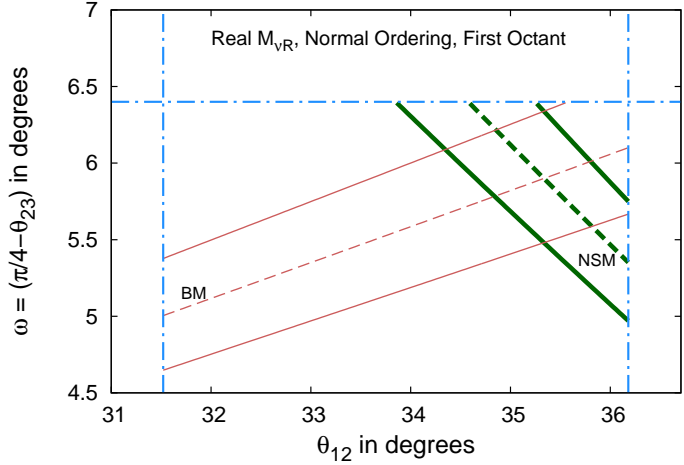}
\end{center}
\caption{\sf  $\omega = (\pi/4 - \theta_{23})$
-vs- $\theta_{12}$ plot for normal ordering. The $3 \sigma$ allowed range 
of $\sin \theta_{13}$ is marked by the solid lines whereas the dashed line
indicates the best-fit value.
Thin pink (thick green) lines denote the BM (NSM) case.
The horizontal and vertical lines represent the data allowed 
3$\sigma$ range. The first octant of $\theta_{23}$ is preferred
since $\omega$ is positive always. 
Although $\omega$ is positive for TBM and GR mixing patterns
its value lies beyond the $3\sigma$ range.
Best-fit values of atmospheric and solar mass splittings are
taken. Inverted ordering is disallowed for $M_{\nu R}$ real.
}
\label{Real} 
\end{figure} 

Using Eqs. (\ref{solspltr}), (\ref{kappa}),  and
(\ref{s13r}) it can be found:
\begin{equation}
\Delta m^2_{solar} =  2~m^- m^{(0)}_1 
~\frac{\sin \theta_{13} \cos\delta \cos \epsilon}{\cos 2\zeta ~\sin
(\epsilon - \theta_{12}^0)}  \;\;. 
\label{solsplr2}
\end{equation}
For real $M_{\nu R}$
inverted ordering is forbidden as can be seen from Eq. (\ref{solsplr2}). 
In order to justify this one can define:
\begin{equation} 
z \equiv m^- m^{(0)}_1/\Delta m^2_{atmos} \;\; {\rm and} \;\;
\tan \xi \equiv m_0/\sqrt{|\Delta m^2_{atmos}|} \;\;,
\label{zxi}
\end{equation}
where $z$ is positive for both the orderings of neutrino masses. 
With the help of Eq.
(\ref{solsplr2}) it can be written as:
\begin{equation}
z = \left(\frac{\Delta m^2_{solar}}{|\Delta m^2_{atmos}|}\right)
\left(\frac{\cos 2\zeta ~\sin (\epsilon - \theta_{12}^0)}{2 \sin
\theta_{13} |\cos\delta| \cos \epsilon}\right) \;\;.
\label{z}
\end{equation} 
From Eq. (\ref{zxi}) it is straightforward to show that:
\begin{eqnarray}
z &=& \sin \xi/(1+ \sin \xi) \;\; {\rm ~i.e.,} \;\; 0 \leq z \leq
\frac{1}{2} \;\; {\rm (for ~normal ~ordering)},\nonumber \\
z &=& 1/(1+ \sin \xi) \;\; {\rm ~i.e.,} \;\; \frac{1}{2} \leq z \leq 1 \;\;
{\rm (for  ~inverted  ~ordering)} \;\;. 
\label{m_0}
\end{eqnarray} 
The lightest neutrino mass $m_0$ has a 
one-to-one correspondence with $z$.
In the quasi-degenerate limit, i.e., $m_0
\rightarrow $ large, $z \rightarrow \frac{1}{2}$ for both orderings.
For real $M_{\nu R}$, $|\cos\delta| = 1$ in  Eq. (\ref{z}).
It simply follows from 
the global fit mass splittings and mixing angles in Sec.
\ref{sec:data} and Table \ref{tlim} that $z\sim 10^{-2}$ or
smaller for all four popular mixing alternatives. 
Thus inverted ordering is forbidden for 
real $M_{\nu R}$.

Using Eqs. (\ref{s13r}) and (\ref{omegar}) the deviation
of the atmospheric mixing angle from maximality is found to be:
\begin{equation}
\tan \omega =  
\frac{\sin \theta_{13}\cos\delta}{\tan (\epsilon -
\theta_{12}^0)} \;\;.
\label{omegar2}
\end{equation}
Eq.
(\ref{omegar}) implies that $\omega$ is positive
always for normal ordering irrespective of the mixing pattern. Thus $\theta_{23}$ is confined only
to the first octant for real
$M_{\nu R}$.
$\epsilon$ can be expressed in terms of $\theta_{12}$ 
using Eqs. (\ref{solangr}) and (\ref{etar}). Thus $\omega$ in
Eq. (\ref{omegar2}) can be expressed as a function of $\theta_{13}$ and
$\theta_{12}$ only.
Fig. \ref{Real} exhibits $\omega$ as a function of $\theta_{12}$
for BM (thin pink lines) and NSM
(thick green lines) alternatives. $\theta_{12}$ and $\omega$
varied within 3$\sigma$ allowed ranges as shown in Sec.
\ref{sec:data}.
The TBM and GR cases are excluded owing as
for the allowed values of $\theta_{12}$ they predict $\theta_{23}$
beyond the 3$\sigma$ range. The 3$\sigma$ limiting values of $\theta_{13}$ are marked
by the solid lines whereas the dashed lines indicate its best-fit value. 
The vertical and horizontal blue dot-dashed lines
denote the 3$\sigma$  experimental
limits of $\theta_{12}$ and $\theta_{23}$. 

With the help of Eq. (\ref{z}), one can translate any allowed point in the $\omega - \theta_{12}$ plane
and the $\theta_{13}$ associated with it to
a value of
$z$ or equivalently $m_0$, when the solar and the atmospheric 
mass splittings are provided. For both the allowed
mixing patterns $m_0$ varies over a very small range.
This range is found to be $2.13$ meV
$\leq m_0 \leq 3.10$ meV ($3.20$ meV $\leq m_0 \leq 4.42$ meV)
for NSM (BM) when both mass splittings and all the three
mixing angles are allowed to vary over their entire 
3$\sigma$ ranges.

The salient features of real $M_{\nu R}$ case are:
\begin{enumerate}
\item Only the normal ordering of neutrino masses is allowed.
\item Only the first octant of $\theta_{23}$ is admissible.
\item Type-I see-saw corrections is unable to make the TBM and GR mixing patterns consistent with the 
allowed ranges of the mixing angles.
\item NSM and BM alternatives can produce solutions
in agreement with the observed neutrino masses and mixing.
The allowed ranges of lightest neutrino mass is very narrow.
\end{enumerate}

\subsection{Complex $M_{\nu R}$}
Real $M_{\nu R}$ has several limitations
viz. inverted ordering and CP-violation is forbidden.
Moreover TBM and GR mixing patterns cannot be
included within the ambit of the model when $M_{\nu R}$ is real.
In order to overcome these constraints the general complex form of 
$M_{\nu R}$ leading to Type-I see-saw contribution $M'$ furnished in
Eq. (\ref{pertmat}) has to be considered. It is worth reminding
ourselves that this choice introduces the complex phases
$\phi_{1,2}$ while $x$ and $y$ can only be positive.

Thus, $M'$ is no longer hermitian. To retain the 
hermitian nature the combination $(M^0 +
M')^\dagger(M^0 + M')$ is considered among which 
$M^{0\dagger} M^0$ and $(M^{0\dagger} M' + M'^\dagger M^0)$
are treated as the leading term and the perturbation at the lowest order
respectively. The unperturbed eigenvalues are given by
$(m^{(0)}_i)^2$ and perturbation matrix is:
\begin{equation}
(M^{0\dagger} M' + M'^\dagger M^0) = 
{m_D^2 \over m_R}
\pmatrix{ 0 & 2 y m^{(0)}_1  \cos\phi_1 & 
y f(\phi_1) \cr
2 y m^{(0)}_1 \cos\phi_1 & \sqrt{2} x m^{(0)}_1  \cos\phi_2 & 
-{x \over\sqrt{2}}  f(\phi_2)\cr
y f^*(\phi_1)& -{x \over\sqrt{2}} f^*(\phi_2)& 
\sqrt{2} x m^{(0)}_3 \cos\phi_2} \;,
\label{pertcmplx1}
\end{equation}
where,
\begin{equation}
f(\varphi) \equiv m^{+} \cos\varphi - i m^{-} \sin\varphi \;\;.
\label{ffn}
\end{equation}
The rest of the procedure is analogous to what was done in case of 
real $M_{\nu R}$ keeping in mind the
discriminating factors of Eq. (\ref{pertcmplx1}).
Now, instead of Eqs.
(\ref{solangr}) and (\ref{etar}) of the real $M_{\nu R}$ case, 
the solar mixing obtained from Eq. (\ref{pertcmplx1}) is given by
\begin{equation}
\theta_{12} = \theta_{12}^0 + \zeta \;\;,\;\; \tan 2\zeta =
2\sqrt2 ~{y \over x} ~{\cos\phi_1\over\cos\phi_2} \;\;,
\label{solangcmplx}
\end{equation}
and
\begin{equation}
\sin \epsilon = \frac{y \cos\phi_1}{\sqrt{y^2 \cos^2\phi_1 + x^2
\cos^2\phi_2/2}}  \;\;,\;\;
\cos \epsilon = \frac{x \cos\phi_2/\sqrt{2}}{\sqrt{y^2 \cos^2\phi_1 + x^2
\cos^2\phi_2/2}}  \;\;,\;\;
\tan \epsilon = \frac{1}{2} \tan 2\zeta \;.
\label{etac}
\end{equation}
\begin{table}[h]
\begin{center}
\begin{tabular}{|c|c|c|c|c|}
\hline
Mixing & \multicolumn{2}{|c|}{Normal Ordering}  & 
\multicolumn{2}{c|}{Inverted Ordering} \\ \cline{2-5} 
Pattern & $\delta$ & $\theta_{23}$& $\delta$
& $\theta_{23}$  \\ 
& quadrant & octant & quadrant & octant \\
\hline
NSM & First/Fourth &
First & Second/Third & Second    \\ \hline
BM, TBM, GR &   Second/Third  &
First  & First/Fourth & Second  \\  \hline
\end{tabular}
\end{center}
\caption{\sf The octant of $\theta_{23}$
and the quadrants of the CP-phase $\delta$
for
different mixing patterns for both orderings of neutrino masses
are exhibited. }
\label{tabCP}
\end{table}
Table \ref{tlim} shows the allowed ranges of $\zeta$ and $\epsilon$
which depend on the mixing patterns. For all mixing alternatives
$\cos \epsilon$ is found to be positive.
Thus from Eq.
(\ref{etac}) $\phi_2$ must 
always lie in the first or fourth quadrants.
For the different mixing patterns the ranges of $\phi_1$
are also given by that of $\epsilon$.
When $\epsilon$ is positive (negative) then from the
first relation contained in  Eq. (\ref{etac}), it is evident that 
$\phi_1$ has
to be in the first or fourth (second or third) quadrants.
Using the results displayed in Table
\ref{tlim} one can infer 
that the first (second) option holds for 
the NSM (BM) patterns. 
In case of TBM and GR, $\epsilon$ varies over 
positive and negative values making both options equally admissible.

Applying degenerate perturbation 
theory the solar mass splitting
attributed completely to the Type-I
see-saw contribution can be obtained from Eq. (\ref{pertcmplx1}): 
\begin{equation}
\Delta m^2_{solar}
= \sqrt{2} m^{(0)}_1 ~\frac{m_D^2 }{m_R} \sqrt{x^2 \cos^2\phi_2
+ 8 y^2 \cos^2 \phi_1} 
= \sqrt{2} m^{(0)}_1 ~\frac{m_D^2 }{m_R} ~\frac{x \cos
\phi_2}{\cos 2\zeta} = \sqrt{2} m^{(0)}_1 ~\frac{m_D^2 }{m_R}
 \frac{2\sqrt{2}y \cos\phi_1}{\sin 2\zeta}
\;.
\label{solsplc1}
\end{equation}
In place of Eq. (\ref{psi3_1}) one gets: 
\begin{equation}
|\psi_3\rangle =
\pmatrix{\kappa_c [\frac{\sin\epsilon}{\cos\phi_1} f(\phi_1) \cos \theta_{12}^0 - 
\frac{\cos\epsilon}{\cos\phi_2} f(\phi_2) \sin \theta_{12}^0] /m^+ \cr 
{1\over \sqrt 2}\{1- \kappa_c [\frac{\sin\epsilon}{\cos\phi_1}
f(\phi_1) \sin \theta_{12}^0 +
\frac{\cos\epsilon}{\cos\phi_2}  f(\phi_2) \cos \theta_{12}^0]/m^+\} \cr 
{1\over \sqrt 2}\{1+ \kappa_c [\frac{\sin\epsilon}{\cos\phi_1}
f(\phi_1) \sin \theta_{12}^0 +
\frac{\cos\epsilon}{\cos\phi_2}  f(\phi_2) \cos \theta_{12}^0]/m^+\}
} \;,
\label{psi3ca}
\end{equation}
where,
\begin{equation}
\kappa_c = \frac{m_D^2}{m_R m^-} ~{\sqrt{y^2 \cos^2\phi_1 + x^2
\cos^2\phi_2/2}} \;,
\end{equation}
Here Eq. (\ref{etac})
and the complex function $f(\phi_{1,2})$ defined in Eq. (\ref{ffn})
have been used. $\kappa_c$ is positive
(negative) for NO (IO).
Comparing Eq. (\ref{psi3ca}) with the
third column of Eq. (\ref{PMNS}) leads to:
\begin{equation}
\sin \theta_{13}\cos\delta = \kappa_c  ~\sin(\epsilon-
\theta_{12}^0) \;, 
\label{cdelcomp}  
\end{equation}
\begin{equation}
\sin \theta_{13}\sin\delta = \kappa_c ~\frac{m^-}{m^+
\cos\phi_1 \cos\phi_2}  
\left[\sin \epsilon \sin\phi_1 \cos \phi_2  \cos \theta_{12}^0
- \cos \epsilon \cos\phi_1 \sin \phi_2  \sin \theta_{12}^0 \right] \;.
\label{sdelcomp}
\end{equation}

\begin{figure}[tbh]%
\begin{center}
{\includegraphics[scale=0.65,angle=0]{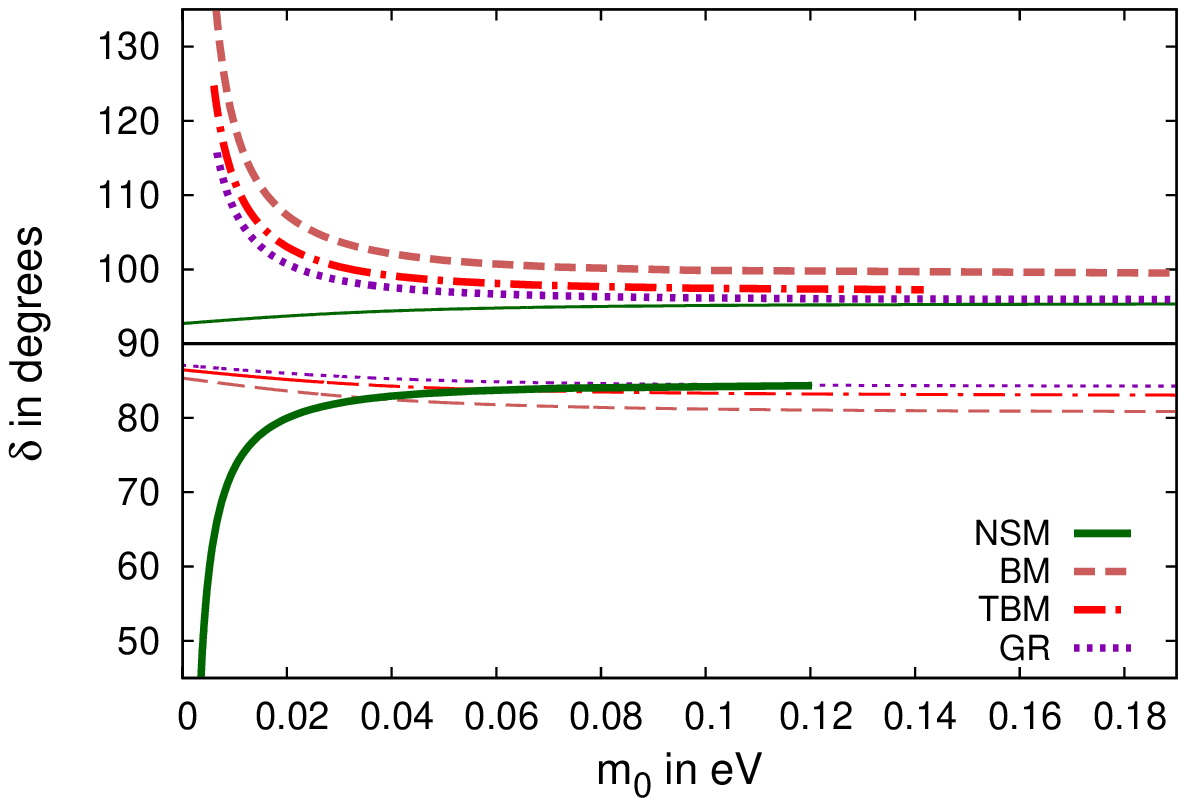}
\hspace*{0.25pt}
\includegraphics[scale=0.65,angle=0]{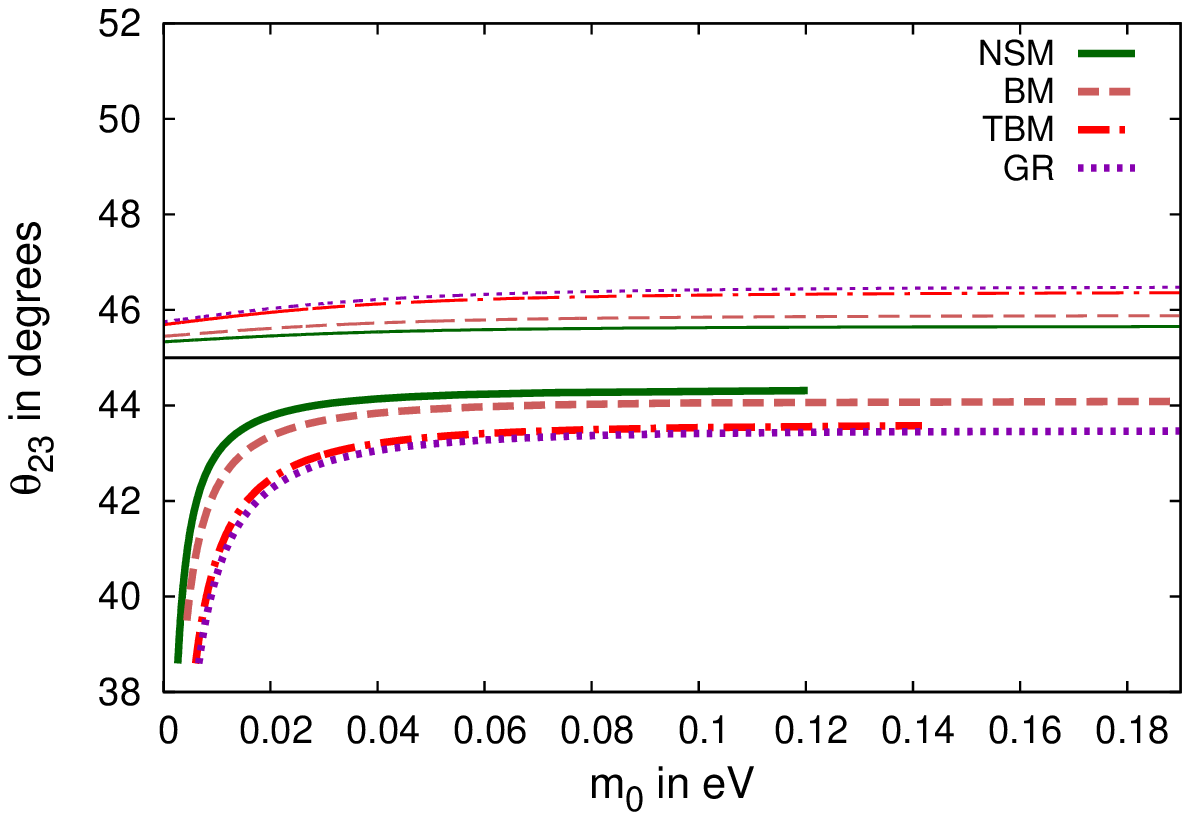}}
\caption{\sf
In the left (right) panel the CP-phase $\delta$ ($\theta_{23}$)
predicted by this model
is plotted as a function of the lightest neutrino
mass $m_0$ for all the four
mixing patterns when the best-fit values of the data are
taken as input. The NSM, BM, TBM and GR mixing alternatives
are represented by the green solid, pink
dashed, red dot-dashed, and violet dotted curves respectively.
Thick (thin) curves of each kind denote NO (IO).}
\label{f:CP} 
\end{center} 
\end{figure} 
From Table \ref{tlim}, it is obvious that $(\epsilon - \theta_{12}^0)$ 
exists in the first (fourth) quadrant for the NSM (BM, TBM, and
GR) mixing pattern. From Eq. (\ref{cdelcomp}) one can immediately conclude
that for NSM (BM, TBM, and GR) case(s) $\delta$ remains in
the first or fourth (second or third) quadrants in case of normal ordering.
$\kappa_c$ changes sign for inverted ordering. Thus the quadrants
get modified accordingly. The different alternatives are furnished
in Table \ref{tabCP}. There are two allowed quadrants of $\delta$ having
$\sin \delta$ of opposite sign for any mixing option and 
ordering of neutrino masses. The sign of the
right-hand-side of Eq. (\ref{sdelcomp}) governs the phases $\phi_{1,2}$ 
which in its turn decides the quadrants CP-phase
$\delta$ out of the two allowed options. As already discussed,
$\phi_2$ can be in either the first or fourth quadrants.
The quadrant of $\phi_1$ depends on the mixing pattern in such
a manner that $\sin \phi_1$ can be of either sign.
Therefore, the phases $\phi_1$ and $\phi_2$ 
can be chosen in a way such that $\sin \delta$ 
can acquire any particular sign.
Thus the two alternate quadrants of $\delta$ for every case in Table
\ref{tabCP} are equally allowed in the model.

The Type-I see-saw perturbative contribution to the atmospheric
mixing angle can be obtained from Eq. (\ref{psi3ca}) as:
\begin{equation}
\tan\omega = \frac{\sin\theta_{13} \cos\delta}{\tan(\epsilon - \theta_{12}^0)}
\;.
\label{th23}
\end{equation}

Let us recall, Eq. (\ref{cdelcomp}) relates $\delta$ and
$(\epsilon- \theta_{12}^0)$ through $\kappa_c$. Thus 
for all mixing alternatives $~\theta_{23}$
always remains in first (second) octant for NO (IO).
This is one of the most important results of the model as shown in
Table \ref{tabCP}.

In the solar splitting expressed in Eq. (\ref{solsplc1}),
the factor of ~$m_D^2/m_R$ can be replaced in terms of $\kappa_c$.
This together with Eq. (\ref{cdelcomp}) gives,
\begin{equation}
\Delta m^2_{solar}
= \frac{2 m^- m_1^{(0)} \sin \theta_{13} \cos\delta \cos\epsilon}
{\sin(\epsilon - \theta_{12}^0)\ cos 2\zeta} 
\;.
\label{solsplc2}
\end{equation}

Predictions of the model can be extracted from Eqs. (\ref{th23}) and
(\ref{solsplc2}). The three mixing angles $\theta_{13}$,
$\theta_{12}$, and  $\theta_{23}$ are taken as inputs.  
Eq. (\ref{th23}) determines a value of the CP-violating phase $\delta$. 
With the help of these and the experimentally observed solar splitting
the combination
$m_1^{(0)} m^-$, or equivalently the variable $z$ can be calculated
using Eq. (\ref{solsplc2}) that fixes the lightest neutrino mass $m_0$.
It may seem that arbitrarily
large values of $m_0$, and hence $m_1^{(0)} m^-$, may be accounted for
by tuning $\cos \delta$ to smaller and smaller values. 
However, this certainly is not the case.
Experimental data necessitate $\omega = (\pi/4 -
\theta_{23})$ to be restricted within observed limits.
As all other
factors have ranges determined experimentally, 
Eq. (\ref{th23}) also puts lower and upper bounds on
$\delta$.
Subsequently, $m_0$ lies within
a fixed range for any mixing pattern.

Fig. \ref{f:CP} contains the CP-phase $\delta$ ($\theta_{23}$)
as a function of the
lightest neutrino mass $m_0$ for
different mixing patterns as predicted by this model 
in the left (right) panel while 
the best-fit values of the various
measured angles and mass splittings are used. 
The NSM, BM, TBM
and GR are depicted by green solid, pink dashed, red
dot-dashed, and violet dotted curves respectively. 
The thick
(thin) curves of each kind indicate NO (IO).
Normal and inverted orderings 
are always associated with the first and second octants of the atmospheric mixing angle 
$\theta_{23}$ respectively. For NSM case $\delta$ lies in the first (second) quadrant
for normal (inverted) ordering, while for the rest of the mixing options 
it is in the second (first) quadrant.
For inverted neutrino mass ordering $|\delta|$ remains close to $\pi/2$
for the complete range of $m_0$. The CP-phase $\delta$ lies near $\pi/2$
for normal ordering for 
$m_0$  larger than around 0.05 eV.

From Table \ref{tabCP} it is evident that if 
$\delta$ is a solution for
some $m_0$ then by properly choosing alternate values of the
phases $\phi_{1,2}$ appearing in $M_{\nu R}$ one can also obtain
a second solution with the phase $-\delta$. 
This mirror set of solutions are not shown in Fig. \ref{f:CP}.
The preliminary data presented by 
the  T2K
\cite{T2K_s3} and NOVA \cite{Nova_s3} collaborations
can be considered as primary hint of normal ordering
associated with $\delta \sim -\pi/2$.
The consistency of this model with these observations
is clearly visible from Fig.
\ref{f:CP} with $\delta \sim
-\pi/2$ favouring $m_0$ in the quasi-degenerate regime, i.e.,
$m_0 \geq \mathcal{O}$(0.05 eV), for normal ordering. 
If this result is determined with better accuracy
in the future analysis then the model will predict 
neutrino masses to be in a range that ongoing experiments
are capable of probing \cite{katrin_s3, numass_s3}.

These interrelationships between the octant of
$\theta_{23}$, the quadrant of the CP-violating phase $\delta$, and the
neutrino mass ordering  provide a clear set of correlations characteristic of 
this $A4$ based model. In the model the corrections
to the three neutrino mixing angles and $\Delta m^2_{solar}$
all have a common origin -- the Type-I see-saw.
As a result these parameters get correlated. 
Such interrelationships are specific to this model.
Although the model has a large number of 
parameters, only this correlated region of the parameter space
allowed by neutrino mass and mixing data leads to testable 
predictions in Table \ref{tabCP}.

\section{Conclusions}

In this paper an $A4$ based see-saw model
for neutrino masses and mixing has been 
proposed. The flavour quantum numbers suitable for the model 
are assigned to the leptons and the scalars. The Lagrangian
is inclusive of all the symmetry conserving terms. No soft 
breaking of symmetry is entertained. 
The Yukawa couplings induce the charged
lepton masses, Dirac and Majorana masses for the
left- and right-handed neutrinos after the symmetry 
is broken spontaneously.
Neutrino masses are produced by a combined effect of 
both Type-I and Type-II see-saw terms present in the 
Lagrangian of which  the former can be
thought of to be a small correction.  
The Type-II see-saw dominant contribution
is associated with the atmospheric mass splitting, no solar splitting,
keeps $\theta_{23} = \pi/4$, and $\theta_{13} = 0$ and 
$\theta_{12}$ can be given any preferred
value. In particular, this model is scrutinized in context of
tribimaximal, bimaximal, golden ratio, and `no
solar mixing' patterns. The contribution of Type-I see-saw
can be treated as a perturbation that generates the solar splitting
and tunes the mixing angles to values in agreement with the global
fits. As a corollary a correlation between
the octants of $\theta_{23}$ and neutrino mass ordering 
followed -- first (second) octant is allowed for
normal (inverted) ordering of neutrino mass.
The model has several testable predictions including that
of the CP-phase $\delta$, relationships
between mixing angles and mass splittings.
Moreover, inverted ordering got associated with near-maximal
CP-phase $\delta$ and arbitrarily small neutrino masses are
allowed.
In case of normal ordering $\delta$ can vary over a
larger range and maximality is accomplished in the quasi-degenerate
regime. The lightest neutrino mass has to be at least a few meV for
this case.


{\bf Acknowledgements:} I acknowledge support from CSIR, India
during the initial stage of this work at University of Calcutta. 
I also thank
Prof. Amitava Raychaudhuri for his valuable insights and comments at 
various stages of this endeavor.

\renewcommand{\thesection}{\Alph{section}} 
\setcounter{section}{0} 
\renewcommand{\theequation}{\thesection.\arabic{equation}}

\setcounter{equation}{0}
\section{Appendix: The group $A4$ }\label{GroupA4}

$A4$ is the even permutation group of four objects having 
12 elements and two generators $S$ and $T$  satisfying the 
property $S^2=T^3=(ST)^3=\mathbb{I}$.
It has four inequivalent irreducible representations viz. one
3 dimensional representation and three 1 dimensional representations namely, $1, 1'$ and
$1''$. These three dimension-1 representations are singlets under $S$ whereas they transform as 
1, $\omega$, and $\omega^2$ respectively under the action of $T$, $\omega$ being a cube root of unity.
Therefore it is apparent that $1'\times 1'' = 1$. 
The pertinent form of the generators $S$ and $T$  
acting on the 3 dimensional representations are given by\footnote{This choice of 
basis has the generator $S$ diagonal. One can equivalently perform an analogous analysis in a basis
in which the generator $T$ is diagonal. Needless to mention that the two bases are related by
some unitary basis transformation.},
\begin{equation}
S=\pmatrix{1 & 0 & 0 \cr 0 & -1 & 0 \cr 0 & 0 & -1}\ \ \ \ {\rm and} \ \ \ \
T=
\pmatrix{0 & 1 & 0 \cr
0 & 0 & 1 \cr
1 & 0 & 0} \;\;.
\label{ST3}
\end{equation}
It is imperative to note the product rule for the three dimensional representation is:
\begin{equation}
3\otimes3=1 \oplus 1' \oplus 1'' \oplus 3 \oplus 3 \;\;.
\label{A43x3}
\end{equation}
When two triplets of $A4$ given by $3_a \equiv {a_i}$ and $3_b \equiv {b_i}$, with
$i=1,2,3$; are combined according to Eq. (\ref{A43x3}), then the resultant triplets 
can be represented by $3_c \equiv {c_i}$
and $3_d \equiv {d_i}$ where,
\begin{eqnarray}
c_i &=& \left(\frac{a_2 b_3 + a_3 b_2}{2},\frac{a_3 b_1 + a_1 b_3}{2},
\frac{a_1 b_2 + a_2 b_1}{2}\right)   \;\;, \;\;{\rm or,}
\;\; c_i \equiv \alpha_{ijk} a_j b_k \;\;, \nonumber \\
d_i &=& \left(\frac{a_2 b_3 - a_3 b_2}{2},\frac{a_3 b_1 - a_1 b_3}{2},
\frac{a_1 b_2 - a_2 b_1}{2}\right)   \;\;, \;\;{\rm or,}
\;\; d_i \equiv \beta_{ijk} a_j b_k \;\;,\;\; (i,j,k, {\rm are ~cyclic})\;\;.
\label{3x3to3}
\end{eqnarray}
and the $1$, $1'$ and $1''$ so obtained can be scripted as: 
\begin{eqnarray}
1 &=& a_1b_1+a_2b_2+a_3b_3 \equiv \rho_{1ij}a_ib_j \;\;,\nonumber \\  
1'&=& a_1b_1+\omega^2a_2b_2+\omega a_3b_3 \equiv \rho_{3ij}a_ib_j
\;\;,  \nonumber \\
1'' &=& a_1b_1+\omega a_2b_2+\omega^2a_3b_3 \equiv \rho_{2ij}a_ib_j \;\;. 
\label{3x3to1}
\end{eqnarray}
The group is studied in extensive details in \cite{A4mr, A4af}.


\setcounter{equation}{0}  
\section{Appendix: Minimization of the scalar potential}\label{V_A4}



Some detailed analysis of the nature of the scalar potential is
presented in this Appendix. The conditions that
have to be satisfied by the parameters of the
potential so that the {\em vev}s acquire the values
considered in the model are extracted. The
conditions so obtained guarantee the potential is locally minimized by those choices.
To confirm if those choices are in concurrence 
with the {\em global} minimum is beyond the scope of
this work\footnote{As an example one can take a look at \cite{gmin1}
where a comparatively simpler scenario consisting of an $A4$ triplet
composed of three $SU(2)_L$ doublet
scalar or in other words an $A4$ symmetric three Higgs doublet model (3HDM) was 
analyzed in terms of the global minimization of the scalar potential.
In \cite{we_3hdm}, it is 
shown that alignment follows as a natural
consequence when the {\em vev}s acquire the configurations
corresponding to those global minima.
Three Higgs doublets symmetric under $A4$ group has been vividly discussed in \cite{Toorop_3hdm}. 
A model for leptons using an $A4$ symmetric 3HDM can be found in \cite{gmin2}.}.

The fields catalogued in Table \ref{tab1s} comprise of scalars 
having lepton numbers as well as $A4$,
$SU(2)_L$, and $U(1)_Y$ charges.
The scalar potential must be of the most general quartic nature
conserving all the symmetries under consideration.
Thus all the terms allowed by the symmetries are included in the discussion
below. Verification of $SU(2)_L$, $U(1)_Y$ and lepton number are familiar exercises.
$A4$ invariance requires elaborate discussion as presented in the following section.

\subsection{$A4$ conserving terms: Notations and general principles}

Let us summarize a few salient features of this model to fix the notations to be followed
for the $A4$-invariant terms. As already noted, the scalar spectrum
has fields transforming as $1, 1',
1'',$ and $3$ under $A4$. One has to consider all the combinations of these
fields up to quartics that can yield $A4$ invariants. The product rules for
$1, 1'$ and $1''$ are easy, but that for the triplets of $A4$ needs to be 
emphasized. If there are two $A4$ triplet fields
$A\equiv(a_1,a_2,a_3)^T$ and $B\equiv(b_1,b_2,b_3)^T$ where $a_i,
b_i$ may possess $SU(2)_L \times U(1)_Y$ transformation properties
that are not considered for the time being in
the immediate course of discussion.  As furnished in Eq. (\ref{A43x3}),
one can combine $A$ and $B$  to obtain
\begin{equation}
3_A\otimes3_B = 1 \oplus 1' \oplus 1'' \oplus 3 \oplus 3  \;\;.
\label{A1}
\end{equation}

For notational simplicity let us denote
the irreducible representations on the right-hand-side
by  $O_1(A, B)$, $O_{2}(A, B)$, $O_{3}(A, B)$, $T_s(A, B)$ and
$T_a(A, B)$, respectively, where, as already noted in Eqs.
(\ref{3x3to3}, \ref{3x3to1})
\begin{eqnarray}
O_{1}(A, B)&\equiv& 1 = a_1 b_1+a_2 b_2+a_3 b_3 \equiv
\rho_{1ij}a_i b_j \;\;,\nonumber \\
O_{2}(A, B)&\equiv& 1'= a_1 b_1+\omega^2 a_2 b_2+\omega a_3 b_3\equiv
\rho_{3ij}a_i b_j \;\;,  \nonumber \\
O_{3}(A, B) &\equiv& 1''= a_1 b_1+\omega a_2 b_2+\omega^2 a_3 b_3
\equiv \rho_{2ij}a_i b_j
\;\;,
\label{A2}
\end{eqnarray}  
and
\begin{eqnarray}
T_s(A, B)&\equiv& 3= \left(\frac{a_2 b_3 + a_3 b_2}{2}, \ \
\frac{a_3 b_1 + a_1 b_3}{2}, \ \
\frac{a_1 b_2 + a_2 b_1}{2}\right)^T
  \;\;,\nonumber\\
T_a(A, B)&\equiv& 3= \left(\frac{a_2 b_3 - a_3 b_2}{2}, \ \ \frac{a_3 b_1 - a_1 b_3}{2}, \ \ 
\frac{a_1 b_2 - a_2 b_1}{2}\right)^T \;\;.
\label{A3}
\end{eqnarray}
It is worth noticing that $O_3(A^\dagger,A) = [O_2(A^\dagger,A)]^\dagger$ and
$T_a(A, A) = 0$.

The scalar potential can be formulated implementing this notation and keeping in mind
that the scalar sector of this model is devoid of any field which is invariant
under all the symmetries under consideration. Therefore the scalar potential
will contain terms of the following kind (only $A4$ properties are exhibited):
\begin{enumerate} 
\item Quadratic: $W^\dagger W$,
\item Cubic: $X_{i} X'_{j} X''_{k}, X_{i} X_{j} X_{k}$,
$X'_{i} X'_{j} X'_{k},$ $X''_{i} X''_{j} X''_{k}$,
$O_{1}(Y_i,Y_j) X_k$, $O_{2}(Y_i,Y_j) X''_k$, 
$O_{3}(Y_i,Y_j) X'_k$,
\item Quartic: $(W_i^\dagger W_i)(W_j^\dagger W_j), ~(X_i X_j)
(X_k X_l), ~(X_i X_j) (X'_k X''_l), ~(X'_i X''_j)
(X'_k X''_l), ~(X'_i X'_j)(X'_k X_l), ~(X''_i X''_j)(X''_k X_l)$,\\
$O_{1}(Y_i,Y_j) X_k X_l, ~O_{1}(Y_i,Y_j) X'_k X''_l,
~O_{2}(Y_i,Y_j) X'_k X'_l, ~O_{2}(Y_i,Y_j) X_k X''_l$,\\
$O_{3}(Y_i,Y_j) X''_k X''_l, ~O_{3}(Y_i,Y_j) X_k X'_l$,\\
$O_{1}(Y_i,Y_j) O_{1}(Y_k,Y_l), O_{2}(Y_i,Y_j)^\dagger
O_{2}(Y_k,Y_l), O_{3}(Y_i,Y_j)^\dagger O_{3}(Y_k,Y_l),
O_{2}(Y_i,Y_j) O_{3}(Y_k,Y_l)$,\\
 $O_{1}(T_s(Y_i,Y_j), T_s(Y_k,Y_l))$,
$O_{1}(T_s(Y_i,Y_j), T_a(Y_k,Y_l))$,
$O_{1}(T_a(Y_i,Y_j), T_a(Y_k,Y_l))$.\\
$O_1(T_s(Y_i,Y_j), Y_k) X_l$, $O_{2}(T_s(Y_i,Y_j), Y_k) X''_l$,
$O_{3}(T_s(Y_i,Y_j), Y_k) X'_l$, \\
$O_1(T_a(Y_i,Y_j), Y_k) X_l$, $O_{2}(T_a(Y_i,Y_j), Y_k) X''_l$,
$O_{3}(T_a(Y_i,Y_j), Y_k) X'_l$ .
\end{enumerate}

Here $W$ is any field, $X$, $X'$, and $X''$ represent
generic fields transforming as $1$, $1'$, and $1''$ under $A4$
while $Y$ happens to be generic $A4$ triplet field.
The invariants constructed by using $X^{\dagger}$,
$X'^{\dagger}$, $X''^{\dagger}$, and $Y^{\dagger}$ are not listed separately. 

Owing to the large number of scalars in the model  --
e.g., $SU(2)_L$ singlets, doublets, and triplets -- the scalar potential
consists of many terms.
In order to simplify the discussion, cubic terms in the fields are excluded
and all the couplings are taken to be real. 
The antisymmetric triplet arising from the combination of two $A4$ triplets i.e.,
the terms denoted by $T_a$ in Eq. (\ref{A3})
are not included in the potential throughout for ease of calculation.
The potential is studied
piece-wise: (a) consisting of terms that arise from combination of fields
belonging to same $SU(2)_L$ sector, and (b) comprising of terms obtained by 
combining scalars of different $SU(2)_L$
sectors. The {\em vev} of the $SU(2)_L$ singlets giving rise to the 
right-handed neutrino mass are larger than the {\em vev} of the other
scalar fields. Thus in the latter category the combinations
of $SU(2)_L$ singlets with the doublets and triplets of $SU(2)_L$ 
are considered, whereas, doublet-triplet inter-sector terms are 
dropped owing to the smallness of the triplet {\em vev} responsible for the
left-handed Majorana neutrino mass. Also the electroweak precision measurements
put a stringent bound on the triplet {\em vev} compelling it to be very small.

\subsection{$SU(2)_L$ Singlet Sector:}
The $SU(2)_L$ singlet scalar sector 
consists of three $A4$ triplets
$\hat \Delta^R_p$ with $p=a,b,c$ denoting each one of them.
These three triplets possess identical quantum numbers, their
{\em vev} being the only discriminating criterion.
Also there are three more fields viz. $\Delta_1^R$,
$\Delta_2^R$ and $\Delta_3^R$ transforming as 
$1$, $1'$ and $1''$ under $A4$. From Eq. (\ref{A1})
we can see that two same
$\hat{\Delta}^{R}_p$ triplets can combine to produce several $A4$
irreducible representations. For notational simplicity let us define:
\begin{equation}
O_{1p}^{ss} \equiv O_{1}(\hat{\Delta}^{R\dagger}_p, \hat\Delta^R_p);
~O_{2p}^{ss} \equiv O_{2}(\hat{\Delta}^{R\dagger}_p, \hat\Delta^R_p);
~T_{sp}^{ss} \equiv T_s(\hat{\Delta}^{R}_p, \hat\Delta^R_p)\;\;, \;\;  
(p=a,b,c).
\label{Prod3s}
\end{equation}
Using two different triplets $\hat \Delta^R_p$ and $\hat{\Delta}^R_q$ where $p\ne q$
analogous combinations can be defined:
\begin{equation}
\hat{O}_{1pq}^{ss} \equiv O_{1}(\hat{\Delta}^{R\dagger}_p, \hat\Delta^R_q);
~\hat{O}_{2pq}^{ss} \equiv O_{2}(\hat{\Delta}^{R\dagger}_p, \hat\Delta^R_q);
~\hat{T}_{spq}^{ss} \equiv T_s(\hat{\Delta}^{R}_p, \hat\Delta^R_q)\;\;, \;\; 
(p,q=a,b,c \ \ {\rm and} \ \ p\ne q).
\label{Prod3sp}
\end{equation}
Generically, it is convenient to use $\widetilde{O}_{ip}$ or
$\widetilde{T}_{sp}$ if the second triplet in the
argument is
replaced by its hermitian conjugate.  As an example,
\begin{equation}
~\widetilde{O}_{1p}^{ss} \equiv O_{1}(\hat{\Delta}^{R\dagger}_p,\hat\Delta^{R\dagger}_p),
~\widetilde{O}_{2p}^{ss} \equiv O_{2}(\hat{\Delta}^{R\dagger}_p,\hat\Delta^{R\dagger}_p),
~\widetilde{O}_{3p}^{ss} \equiv O_{3}(\hat{\Delta}^{R\dagger}_p,
\hat\Delta^{R\dagger}_p) ~{\rm and}
~\widetilde{T}_{sp}^{ss} \equiv T_{s}(\hat{\Delta}^{R}_p, \hat\Delta^{R\dagger}_p),
\end{equation}
One can also consider:
\begin{equation}
~\widetilde{O}_{1pq}^{ss} \equiv O_{1}(\hat{\Delta}^{R\dagger}_p,\hat\Delta^{R\dagger}_q),
~\widetilde{O}_{2pq}^{ss} \equiv O_{2}(\hat{\Delta}^{R\dagger}_p,\hat\Delta^{R\dagger}_q),
~\widetilde{O}_{3pq}^{ss} \equiv O_{3}(\hat{\Delta}^{R\dagger}_p,\hat\Delta^{R\dagger}_q). 
\end{equation}
Also the following combinations are required:
\begin{equation}
~\mathscr{O}_{1p}^{ss} \equiv O_{1}(\hat{\Delta}^{R}_p,
{T}_{sp}^{ss\dagger}),
~\mathscr{O}_{2p}^{ss} \equiv O_{2}(\hat{\Delta}^{R}_p,
{T}_{sp}^{ss\dagger}),
~\mathscr{O}_{3p}^{ss} \equiv O_{3}(\hat{\Delta}^{R}_p,
{T}_{sp}^{ss\dagger})\;\;, \;\;  
(p=a,b,c).
\end{equation}
The $A4$ singlets $\Delta_i^R$ ($i=1,2,3$) can be combined to yield 
\begin{equation}
Q_{i}^{ss} \equiv \Delta_i^{R\dagger} \Delta_i^R \;\;, \;\;  
(i=1,2,3).
\end{equation}
Needless to mention that such terms are singlets of all the symmetries under consideration.

Having devised the essential notations one can write the 
most general scalar potential for the $SU(2)_L$ singlet sector of this model as: 
\begin{eqnarray}
V_{singlet}&=& \sum_{i=1}^3 m_{\Delta_i^R}^2 Q_i^{ss} + \sum_{p=a}^c m_{\hat\Delta^R_p}^2 O_{1p}^{ss} 
+ \left[\sum_{{p\ne q};\,{p,q=a}}^c m_{\hat\Delta^R_{pq}}^2 \hat{O}_{1pq}^{ss} \ \ 
+ \ \ {\rm all} \ \ {\rm possible} \ \ {\rm permutations} \right]
 \nonumber \\
&+& {1 \over 2} \sum_{i=1}^3\lambda^s_{1i}\left[Q_i^{ss}\right]^2 
+{1 \over 2}\sum_{k<j;\, k\ne j;\, k=1}^2 \sum_{j=2}^3 \lambda^s_{2jk} \left\{ Q_j^{ss}Q_k^{ss}\right  \}
 \nonumber \\
&+& {1 \over 2}\sum_{p=a}^c\lambda_{3p}^s\left\{ [O_{1p}^{ss}]^2 +
 (O_{2p}^{ss})^\dagger O_{2p}^{ss} 
+ O_{1p}({T_{sp}^{ss}}, T_{sp}^{ss\dagger})  \right  \}
\nonumber \\ 
&+& \sum_{{p\ne q};\,{p,q=a}}^c \lambda^s_{3pq} \left\{ [\hat{O}_{1pq}^{ss}]^2 +
 (\hat{O}_{2pq}^{ss})^\dagger \hat{O}_{2pq}^{ss} + h.c.\right  \}
+{1 \over 2}\sum_{{p\ne q};\,{p,q=a}}^c\widetilde{\lambda}^s_{3pq}\left\{(\hat{O}_{1pq}^{ss})^\dagger\hat{O}_{1pq}^{ss}+ O_{1}({\hat{T}_{spq}^{ss}}, \hat{T}_{spq}^{ss\dagger})  \right  \}\nonumber \\ 
&+& 
\sum_{p=a}^c \sum_{i=1}^3 \left[{1 \over 2}\lambda^s_{4ip}\left(Q_i^{ss} O_{1p}^{ss} \right)\right]
+\sum_{i=1}^3\sum_{{p\ne q};\,{p,q=a}}^c\lambda^s_{4ipq}\left[\left(Q_i^{ss} \hat{O}_{1pq}^{ss} \right) +h.c.\right]
+{1 \over 2}\sum_{p=a}^c\lambda^s_{5p} \left( \mathscr{O}_{1p}^{ss} \Delta_1^R + ~{\rm h.c.}\right)
\nonumber \\
&+& \sum_{{p\ne q};\,{p,q=a}}^c\lambda^s_{5pq}
\left[
\left\{ 
\left(\Delta_1^R O_1(\hat{\Delta}_p^R, \hat{T}_{spq}^{ss\dagger}\right) +
\left(\Delta_1^R O_1(\hat{\Delta}_q^R, \hat{T}_{spq}^{ss\dagger}\right) 
 \right  \} +h.c.\right]
\nonumber \\
&+&{1 \over 2}\sum_{p=a}^c\lambda^s_{6p} \left( \mathscr{O}_{3p}^{ss} \Delta_2^R + ~{\rm h.c.}\right)
+\sum_{{p\ne q};\,{p,q=a}}^c\lambda^s_{6pq}
\left[
\left\{ 
\left(\Delta_2^R O_3(\hat{\Delta}_p^R, \hat{T}_{spq}^{ss\dagger}\right) +
\left(\Delta_2^R O_3(\hat{\Delta}_q^R, \hat{T}_{spq}^{ss\dagger}\right) 
 \right  \} +h.c.\right]
\nonumber \\
&+&{1 \over 2}\sum_{p=a}^c\lambda^s_{7p} \left( \mathscr{O}_{2p}^{ss} \Delta_3^R + ~{\rm h.c.}\right)
+\sum_{{p\ne q};\,{p,q=a}}^c\lambda^s_{7pq}
\left[
\left\{ 
\left(\Delta_3^R O_2(\hat{\Delta}_p^R, \hat{T}_{spq}^{ss\dagger}\right) +
\left(\Delta_3^R O_2(\hat{\Delta}_q^R, \hat{T}_{spq}^{ss\dagger}\right) 
 \right  \} +h.c.\right]\nonumber \\
&+&
\sum_{p=a}^c\sum_{i=1}^3 \lambda^s_{8ip}\left({\Delta_i^R}^2 \widetilde{O}_{ip} +h.c.\right)
+\sum_{{p\ne q};\,{p,q=a}}^c\sum_{i=1}^3 \lambda^s_{8ipq}\left({\Delta_i^R}^2 \widetilde{O}_{ipq}+h.c.\right)
\nonumber \\
&+&\sum_{p=a}^c \left[\lambda^s_{91p}\Delta_2^R\Delta_3^R\widetilde{O}_{1p}
+\lambda^s_{92p}\Delta_1^R\Delta_3^R\widetilde{O}_{2p}+\lambda^s_{93p}\Delta_1^R\Delta_2^R\widetilde{O}_{3p}
+h.c.\right]
\nonumber \\
&+& \sum_{{p\ne q};\,{p,q=a}}^c\left[\lambda^s_{91pq}\Delta_2^R\Delta_3^R\widetilde{O}_{1pq}
+\lambda^s_{92pq}\Delta_1^R\Delta_3^R\widetilde{O}_{2pq}+\lambda^s_{93pq}\Delta_1^R\Delta_2^R\widetilde{O}_{3pq}
+h.c.\right]
.
\nonumber\\
 \;\;.
\label{Vs}
\end{eqnarray}
Here $\lambda^s_{3p}$, $\lambda^s_{3pq}$ and $\widetilde{\lambda}^s_{3pq}$ are taken as the 
common coefficient of the different $A4$ invariants generated
by combining two $\hat\Delta^R$ and two $(\hat{\Delta}^R)^\dagger$
fields. Similar policy will be adopted for the fields with
other $SU(2)_L$ properties.

\subsection{$SU(2)_L$ Doublet Sector:}

The $SU(2)_L$ doublet scalar precinct consists of the two fields 
$\Phi$ and $\eta$ transforming as $3$ and $1$ of $A4$ 
respectively. Opposite hypercharges are 
assigned to $\Phi$ and $\eta$. 
The $A4$ triplet $\Phi$ combinations are denoted as:
\begin{equation}
O_{1}^{dd} \equiv O_{1}(\Phi^\dagger,\Phi);
~O_{2}^{dd} \equiv O_{2}(\Phi^\dagger,\Phi);
~T_s^{dd} \equiv T_s(\Phi,\Phi),
\end{equation}
and that of the $A4$ singlet $\eta$ are:
\begin{equation}
Q_{\eta}^{dd} \equiv \eta^{\dagger} \eta\;\;.
\end{equation}

The potential for the $SU(2)_L$ doublet sector is given by:
\begin{eqnarray}
V_{doublet}&=& m_\eta^2 Q_{\eta}^{dd}  + m_\Phi^2 O_{1}^{dd} 
+ {1 \over 2} \lambda^d_1\left[Q_{\eta}^{dd}\right]^2
+{1 \over 2} \lambda^d_2 \left\{[O_{1}^{dd}]^2 + 
\{O_{2}^{dd}\}^\dagger O_{2}^{dd} \right.
\nonumber \\ 
&+& \left.
O_{1}({T_s^{dd}}, T_s^{dd\dagger})  \right \}
+{1 \over 2} \lambda^d_3\left[Q_{\eta}^{dd} O_{1}^{dd}\right].
\label{Vd}
\end{eqnarray}

\subsection{$SU(2)_L$ Triplet Sector:}

\vskip 10pt
The $SU(2)_L$ triplet sector comprises of five fields. There are two $A4$ triplets 
$\hat{\Delta}^L_a$ and $\hat{\Delta}^L_b$ together with the
fields the $\Delta^L_1$, $\Delta^L_2$ and $\Delta^L_3$
transforming as $1$, $1'$, $1''$ of $A4$ respectively.
\vskip 1pt
It is useful to define:
\begin{equation}
O_{1n}^{tt} \equiv O_{1}(\hat{\Delta}^{L\dagger}_n, \hat\Delta^L_n);
~O_{2n}^{tt} \equiv O_{2}(\hat{\Delta}^{L\dagger}_n, \hat\Delta^L_n);
~T_{sn}^{tt} \equiv T_s(\hat{\Delta}^{L}_n, \hat\Delta^L_n)
\;\;, \;\;(n=a,b),
\end{equation}
\begin{equation}
\hat{O}_{1nl}^{tt} \equiv O_{1}(\hat{\Delta}^{L\dagger}_n, \hat\Delta^L_l);
~\hat{O}_{2nl}^{tt} \equiv O_{2}(\hat{\Delta}^{L\dagger}_n, \hat\Delta^L_l);
~\hat{O}_{3nl}^{tt} \equiv O_{3}(\hat{\Delta}^{L\dagger}_n, \hat\Delta^L_l);
~\hat{T}_{snl}^{tt} \equiv T_s(\hat{\Delta}^{L}_n, \hat\Delta^L_l),
\;\;(n,l= a,b \ \ {\rm and}  \ \ n\ne l),
\end{equation}
\begin{equation}
Q_{i}^{tt} \equiv \Delta_i^{L\dagger} \Delta_i^L \;\;, \;\; (i=1,2,3),
\end{equation}
and
\begin{equation}
~\mathscr{O}_{\gamma n}^{tt} \equiv O_{\gamma}(\hat{\Delta}^{L}_n,
T_{sn}^{tt\dagger}); 
~\mathscr{O}_{\gamma nl}^{tt} \equiv O_{\gamma}(\hat{\Delta}^{L}_n,
\hat{T}_{sl}^{tt\dagger}), \;\; (\gamma= 1,2,3)\;\; {\rm and}  \ \ (n,l= a,b),
\end{equation}
\begin{equation}
~\widetilde{O}^{tt}_{jn}\equiv O_{j}(\hat{\Delta}^{L\dagger}_n,\hat{\Delta}^{L\dagger}_n);
~\widetilde{O}^{tt}_{jnl}\equiv O_{j}(\hat{\Delta}^{L\dagger}_n,\hat{\Delta}^{L\dagger}_l),
 \;\; (j= 1,2,3)\;\; {\rm and}  \ \ (n,l= a,b \ \ {\rm and}  \ \ n\ne l).
\end{equation}

The scalar potential for this sector:
\begin{eqnarray}
V_{triplet}&=& \sum_{i =1}^3 {m^2_{\Delta^L_i}} ~Q_{i}^{tt}
+\sum_{n=a}^b m_{\hat\Delta^L_n}^2 ~O_{1n}^{tt}  
+\left(\sum_{{n\ne l;\, n,l=a}}^b m_{\hat\Delta^L_{nl}}^2~\hat{O}_{1nl}^{tt}
+\ \ {\rm all} \ \ {\rm possible} \ \ {\rm permutations} \right)
 \nonumber \\
&+&{1 \over 2} \sum_{i =1}^3 \lambda^t_{1_i}
\left[Q_{i}^{tt}\right]^2
+{1 \over 2} \sum_{k < j,\, k =1}^2 \sum_{j =2}^3 
\lambda^t_{2jk} Q_{j}^{tt}Q_{k}^{tt}
+{1 \over 2}\sum_{n=a}^b \lambda^t_{3n} \left \{[O_{1n}^{tt}]^2  +
\{O_{2n}^{tt}\}^\dagger O_{2n}^{tt}  +
O_{1}({T_{sn}^{tt}}, T_{sn}^{tt\dagger}) \right \}\nonumber\\
&+&
{1 \over 2}\sum_{{n\ne l;\, n,l=a}}^b\lambda^t_{3nl} \left \{[\hat O_{1nl}^{tt}]^2  +
\{\hat{O}_{2nl}^{tt}\}^\dagger \hat{O}_{2nl}^{tt} +h.c. \right \}
+{1 \over 2}\sum_{{n\ne l;\, n,l=a}}^b\widetilde{\lambda}^t_{3nl}
\left \{[\hat O_{1nl}^{tt}]^\dagger\hat O_{1nl}^{tt}]+O_1(\hat{T}_{sn}^{tt},\hat{T}_{sn}^{tt\dagger} )
\right \}
\nonumber\\
&+&{1 \over 2}\sum_{j =1}^3\sum_{n =a}^b \lambda^t_{4_{jn}}\left[ \left(\Delta^{L\dagger}_j
\Delta^L_j\right)  O_{1n}^{tt}\right]
+\sum_{j=1}^3\sum_{{n\ne l;\, n,l=a}}^b\lambda^t_{4_{1nl}}\left[ \left(\Delta^{L\dagger}_j
\Delta^L_j \right)\hat{O}_{1n}^{tt} +h.c. \right]
\nonumber\\
&+&{1 \over 2}\sum_{n =a}^b\lambda^t_{5n}\left[\left\{ \Delta_1^L\mathscr{O}^{tt}_{1n}\right \}+h.c.\right]
+ \sum_{{ n,l=a}}^b\lambda^t_{5nl}\left[\left\{\Delta_1^L \mathscr{O}^{tt}_{1nl}\right \}+h.c.\right]
+{1 \over 2}\sum_{n =a}^b\lambda^t_{6n}\left[\left\{ \Delta_2^L\mathscr{O}^{tt}_{3n}\right \}+h.c.\right]
\nonumber\\
&+&\sum_{{n,l=a}}^b\lambda^t_{6nl}\left[\left\{\Delta_2^L \mathscr{O}^{tt}_{3nl}\right \}+h.c.\right]
+{1 \over 2}\sum_{n =a}^b\lambda^t_{7n}\left[\left\{ \Delta_3^L\mathscr{O}^{tt}_{2n}\right \}+h.c.\right]
+ \sum_{{n,l=a}}^b\lambda^t_{7nl}\left[\left\{\Delta_3^L \mathscr{O}^{tt}_{2nl}\right \}+h.c.\right]
\nonumber\\
&+&\sum_{n =a}^b\sum_{j=1}^3\lambda^t_{8jn}\left[ \left({\Delta_j^L}^2\widetilde{O}^{tt}_{jn} \right)+h.c.\right]
+\sum_{{n\ne l;\, n,l=a}}^b\sum_{j=1}^3\lambda^t_{8jnl}\left[ \left({\Delta_j^L}^2\widetilde{O}^{tt}_{jnl} \right)+h.c.\right]\nonumber\\
&+&\sum_{n =a}^b\left[
\left\{\lambda^t_{91n}\left(\Delta_2^L\Delta_3^L\widetilde{O}^{tt}_{1n}\right)\right \}
+\left\{ \lambda^t_{92n}\left(\Delta_1^L\Delta_3^L\widetilde{O}^{tt}_{2n}\right)\right \}
+\left\{ \lambda^t_{93n}\left(\Delta_1^L\Delta_2^L\widetilde{O}^{tt}_{3n}\right)\right \}
+h.c.
\right]
\nonumber\\
&+&\sum_{{n\ne l;\, n,l=a}}^b\left[
\left\{\lambda^t_{91nl}\left(\Delta_2^L\Delta_3^L\widetilde{O}^{tt}_{1nl}\right)\right \}
+\left\{ \lambda^t_{92nl}\left(\Delta_1^L\Delta_3^L\widetilde{O}^{tt}_{2nl}\right)\right \}
+\left\{ \lambda^t_{93nl}\left(\Delta_1^L\Delta_2^L\widetilde{O}^{tt}_{3nl}\right)\right \}
+h.c.
\right]
\nonumber\\
 \;\;.
\label{Vt}
\end{eqnarray}

\subsection{Inter-sector terms in the scalar potential:}
The terms in the scalar potential involving
scalar fields of identical $SU(2)_L$ behavior are already taken into account. 
Apart from them, the scalar potential will also receive contributions from
terms generated by combining scalars of two different $SU(2)_L$ sectors that constitute
the main objective of the following discussion.
In this category the combinations of the $SU(2)_L$ singlet scalars with 
that belonging to either of the
doublet or the triplet sector. The other variety of 
inter-sector terms --
doublet-triplet type -- are
not included. This seems to be a reasonable approximation as 
the {\em vev}s of the singlet fields are the largest.

\subsubsection{Singlet-Doublet inter-sector terms:}

Let us consider the combinations:
\begin{equation}
\widetilde{T}_{sp}^{ss} \equiv T_{s}(\hat{\Delta}^{R}_p,\hat{\Delta}^{R\dagger}_p);\ \ 
\widetilde{T}_{spq}^{ss} \equiv T_{s}(\hat{\Delta}^{R}_p,\hat{\Delta}^{R\dagger}_q)
\;\; {\rm and} \;\; \widetilde{T}_{s}^{dd}
\equiv T_{s}(\Phi, \Phi^{\dagger}), \ \ (p,q=a,b,c \ \ {\rm and} \ \ p\ne q)
\end{equation}
and
\begin{equation}
O_{1sp}^{sd} \equiv O_{1}(\widetilde{T}_s^{dd},\widetilde{T}_{sp}^{ss});\ \
\hat{O}_{1spq}^{sd} \equiv O_{1}(\widetilde{T}_s^{dd},\widetilde{T}_{spq}^{ss});
~\mathscr{O}_{p\gamma}^{sd} \equiv O_{\gamma}(\hat{\Delta}^{R}_p,  
\widetilde{T}_s^{dd})\ \ , \ \ (\gamma=1,2,3) \ \ {\rm and} \ \ (p,q=a,b,c \ \ {\rm with} \ \ p\ne q).
\end{equation}

Using this notations:
\begin{eqnarray}
V_{sd}&=& {1 \over 2} \sum_{i=1}^3\left[\lambda^{sd}_{1i}\left(Q_i^{ss} Q_{\eta}^{dd} \right)
+ \left(\lambda^{sd}_{2i} Q_i^{ss} O_{1}^{dd}\right)\right] 
+ {1 \over 2}\sum_{p=a}^c \lambda^{sd}_{3p}\left[Q_{\eta}^{dd} O_{1p}^{ss}\right]
+{1 \over 2}\sum_{{p\ne q};\,{p,q=a}}^c\left[Q_{\eta}^{dd}\hat{O}^{ss}_{1pq}\right]
\nonumber \\ 
&+&\sum_{p=a}^c\left[
\lambda^{sd}_{4p}\left(\{\mathscr{O}_{1p}^{sd}\}\Delta_1^R
+ h.c. \right)
+\lambda^{sd}_{5p}\left(\{\mathscr{O}_{2p}^{sd}\}\Delta_2^R
+ h.c. \right)
+\lambda^{sd}_{6p}\left(\{\mathscr{O}_{3p}^{sd}\}\Delta_3^R
+ h.c. \right)
\right]
\nonumber \\ 
&+& {1 \over 2} \sum_{p=a}^c\lambda^{sd}_{7p}\left[O_{1}^{dd}O_{1p}^{ss} +
\{O_{2p}^{ss}\}^\dagger O_{2}^{dd} + \{O_{2}^{dd}\}^\dagger
O_{2p}^{ss} +O_{1sp}^{sd}\right]
\nonumber \\ 
&+& {1 \over 2}\sum_{{p\ne q};\,{p,q=a}}^c\lambda^{sd}_{7pq}\left[O_{1}^{dd}\hat{O}_{1pq}^{ss} +
\{\hat{O}_{2pq}^{ss}\}^\dagger O_{2}^{dd} + \{O_{2}^{dd}\}^\dagger
\hat{O}_{2pq}^{ss} +\hat{O}_{1spq}^{sd}\right]
.
\label{Vsd}
\end{eqnarray}
In the last two terms a simplifying assumption of using a common couplings
$\lambda_{7p}^{sd}$ and $\lambda_{7pq}^{sd}$ for the terms in the scalar potential that are generated from various combinations of $(\Phi^\dagger \Phi)
(\hat \Delta^{R\dagger} \hat \Delta^R)$, all four of the fields involved being
triplets of $A4$.

\subsubsection{Singlet-Triplet inter-sector terms:}

In this case the following combinations comes into play:
\begin{eqnarray}
\widetilde{T}^{tt}_{sn} &\equiv& T_s (\hat{\Delta}^L_n,\hat{\Delta}^{L\dagger}_n ); \ \ 
\widetilde{T}^{tt}_{snl} \equiv T_s (\hat{\Delta}^L_n,\hat{\Delta}^{L\dagger}_l ); \ \ 
O_{1snp}^{ts}\equiv O_1(\widetilde{T}^{tt}_{sn}, \widetilde{T}^{ss}_{sp});\ \ 
\hat{O}_{1snpq}^{ts} \equiv O_{1}(\widetilde{T}_{sn}^{tt},\widetilde{T}_{spq}^{ss}); \ \
\nonumber\\
\hat{O}_{1snlp}^{ts} &\equiv& O_{1}(\widetilde{T}_{snl}^{tt},\widetilde{T}_{sp}^{ss}); \ \
\hat{O}_{1snlpq}^{ts} \equiv O_{1}(\widetilde{T}_{snl}^{tt},\widetilde{T}_{spq}^{ss}); \ \
O_{\gamma n p}^{ts}\equiv O_{\gamma}(\hat{\Delta}^{R\dagger}_p,\hat\Delta^L_n);  \ \ 
\widetilde{O}_{\gamma np}^{ts}\equiv O_\gamma(\hat{\Delta}^R_p, \hat{\Delta}^L_n);
\nonumber\\
 \mathscr{O}_{\gamma np}^{ts}  &\equiv&
O_{\gamma}(\widetilde{T}_{sp}^{ss},\hat\Delta^L_n);\ \ 
\mathscr{\widetilde{O}}_{\gamma np}^{ts}  \equiv
O_{\gamma}(\widetilde{T}_{sn}^{tt},\hat\Delta^R_p);\ \ 
\hat{\mathscr{O}}_{\gamma npq}^{ts}  \equiv
O_{\gamma}(\widetilde{T}_{spq}^{ss},\hat\Delta^L_n);\ \ 
\mathscr{\widetilde{O}}_{\gamma nlp}^{ts}  \equiv
O_{\gamma}(\widetilde{T}_{snl}^{tt},\hat\Delta^R_p).
\nonumber\\
\end{eqnarray}
where $\;\;(\gamma = 1,2,3) ;
 \ \ (p,q = a,b,c) \ \ {\rm and}  \ \ (n,l = a,b)$. Needless to mention $p\ne q$ and $n\ne l$. 

Following the convention introduced already:
\begin{equation}
O_{\gamma np}^{ts}  \equiv
O_{\gamma }(\hat{\Delta}^{R\dagger}_p,\hat\Delta^{L}_n); \ \ 
\widetilde{O}_{\gamma np}^{ts}  \equiv
O_{\gamma }(\hat{\Delta}^{R}_p,\hat\Delta^{L}_n)
,\;(\gamma = 1,2,3) ; \ \ (p = a,b,c) \ \ {\rm and}  \ \ (n = a,b).
\end{equation}

The inter-sector potential for this case is given by:
\begin{eqnarray}
V_{ts}&=& {1 \over 2} \sum_{i =1}^3 \sum_{j =1}^3\lambda^{ts}_{1ij}
\left[Q_i^{ss}  Q_{j}^{tt} \right]
+ {1 \over 2} \sum_{j =1}^3\sum_{n =a}^b\lambda^{ts}_{2jn} \left[\left(Q_j^{ss} O_{1n}^{tt}\right)
+h.c.\right]
+ {1 \over 2} \sum_{j =1}^3\sum_{{n\ne l;\, n,l=a}}^b\lambda^{ts}_{2jnl} \left[\left(Q_j^{ss} \hat{O}_{1nl}^{tt}\right)
+h.c.\right]
\nonumber \\ 
&+&{1 \over 2} \sum_{i =1}^3 \sum_{p =a}^c \lambda^{ts}_{3ip}
\left[Q_{i}^{tt} O_{1p}^{ss} \right]
+\sum_{i =1}^3 \sum_{{p\ne q};\,{p,q=a}}^c \lambda^{ts}_{3ipq}
\left[Q_{i}^{tt} \hat{O}_{1pq}^{ss} \right]
\nonumber \\ 
&+&{1 \over 2} \sum_{p =a}^c
\sum_{n =a}^b\lambda^{ts}_{41nnpp}
\left[O_{1n}^{tt}O_{1p}^{ss}+\{O_{2p}^{ss}\}^\dagger
O_{2n}^{tt}
+\{O_{2n}^{tt}\}^\dagger O_{2p}^{ss}+O_{1snp}^{ts}  \right]
\nonumber \\ 
&+&{1 \over 2} \sum_{{p\ne q};\,{p,q=a}}^c
\sum_{n =a}^b\lambda^{ts}_{42nnpq}
\left[O_{1n}^{tt}\hat{O}_{1pq}^{ss}+\{\hat{O}_{2pq}^{ss}\}^\dagger
O_{2n}^{tt}
+\{O_{2n}^{tt}\}^\dagger \hat{O}_{2pq}^{ss}+\hat{O}_{1snpq}^{ts}  \right]
\nonumber \\ 
&+&{1 \over 2} \sum_{p =a}^c
\sum_{{n\ne l;\, n,l=a}}^b
\lambda^{ts}_{43nlpp}
\left[\hat{O}_{1nl}^{tt}O_{1p}^{ss}+\{O_{2p}^{ss}\}^\dagger
\hat{O}_{2nl}^{tt}
+\{\hat{O}_{2nl}^{tt}\}^\dagger O_{2p}^{ss}+\hat{O}_{1snlp}^{ts}  \right]
\nonumber \\ 
&+&{1 \over 2} \sum_{{p\ne q};\,{p,q=a}}^c
\sum_{{n\ne l;\, n,l=a}}^b
\lambda^{ts}_{44nlpq}
\left[\hat{O}_{1nl}^{tt}\hat{O}_{1pq}^{ss}+\{\hat{O}_{2pq}^{ss}\}^\dagger
\hat{O}_{2nl}^{tt}
+\{\hat{O}_{2nl}^{tt}\}^\dagger \hat{O}_{2pq}^{ss}+\hat{O}_{1snlpq}^{ts}  \right]
\nonumber \\ 
&+&\sum_{i =1}^3 \sum_{p =a}^c
\sum_{n =a}^b\lambda^{ts}_{5ippn}\left(
\mathscr{O}_{inp}^{ts}{\Delta^L_i}^\dagger + h.c.\right) 
+\sum_{i =1}^3 \sum_{{p\ne q};\,{p,q=a}}^c
\sum_{n =a}^b\lambda^{ts}_{5ipqn}\left(
\hat{\mathscr{O}}_{inpq}^{ts}{\Delta^L_i}^\dagger + h.c.\right)
\nonumber \\ 
&+&\sum_{i =1}^3 \sum_{p =a}^c
\sum_{n =a}^b\lambda^{ts}_{6innp}\left(
\widetilde{\mathscr{O}}_{inp}^{ts}{\Delta^R_i}^\dagger + h.c.\right) 
+\sum_{i =1}^3 \sum_{p =a}^c\sum_{{n\ne l};\,{n,l=a}}^b
\lambda^{ts}_{6inlp}\left(
\widetilde{\mathscr{O}}_{inlp}^{ts}{\Delta^R_i}^\dagger + h.c.\right)
\nonumber \\  
&+&\sum_{p =a}^c\sum_{n =a}^b\left[\lambda^{ts}_7 O_{1np}^{ts}\left(
\Delta_1^{L\dagger}\Delta_1^R+\Delta_2^{L\dagger}\Delta_2^R+\Delta_3^{L\dagger}\Delta_3^R
\right)+h.c.
\right]
\nonumber \\  
&+&\sum_{p =a}^c\sum_{n =a}^b\left[\lambda^{ts}_8 O_{2np}^{ts}\left(
\Delta_1^{L\dagger}\Delta_3^R+\Delta_2^{L\dagger}\Delta_1^R+\Delta_3^{L\dagger}\Delta_2^R
\right)+h.c.
\right] 
\nonumber \\  
&+&\sum_{p =a}^c\sum_{n =a}^b\left[\lambda^{ts}_9 O_{3np}^{ts}\left(
\Delta_3^{L\dagger}\Delta_1^R+\Delta_1^{L\dagger}\Delta_2^R+\Delta_2^{L\dagger}\Delta_3^R
\right)+h.c.
\right]  
\nonumber \\  
&+&\sum_{p =a}^c\sum_{n =a}^b\left[\lambda^{ts}_{10} \widetilde{O}_{3np}^{ts}\left(
\Delta_3^{L\dagger}\Delta_1^{R\dagger}+\Delta_1^{L\dagger}\Delta_3^{R\dagger}
+\Delta_2^{L\dagger}\Delta_2^{R\dagger}
\right)+h.c.
\right] \nonumber \\  
&+&\sum_{p =a}^c\sum_{n =a}^b\left[\lambda^{ts}_{11} \widetilde{O}_{2np}^{ts}\left(
\Delta_2^{L\dagger}\Delta_1^{R\dagger}+\Delta_1^{L\dagger}\Delta_2^{R\dagger}
+\Delta_3^{L\dagger}\Delta_3^{R\dagger}
\right)+h.c.
\right] \nonumber \\  
&+&\sum_{p =a}^c\sum_{n =a}^b\left[\lambda^{ts}_{12} \widetilde{O}_{1np}^{ts}\left(
\Delta_1^{L\dagger}\Delta_1^{R\dagger}+\Delta_3^{L\dagger}\Delta_2^{R\dagger}
+\Delta_2^{L\dagger}\Delta_3^{R\dagger}
\right)+h.c.
\right] 
\;\;.
\label{Vts}
\end{eqnarray}
It must be noted that while writing the last $\lambda^{ts}_{7-12}$ terms the different couplings 
corresponding to the combinations of $O_{inp}^{ts}$ with $(\Delta^{L\dagger}_i\Delta^R_j)$
and $\widetilde{O}_{inp}^{ts}$ with $(\Delta^{L\dagger}_i\Delta^{R\dagger}_j)$ are set to be equal.

\subsection{The conditions for minimization:}
With the scalar potential in hand it is necessary to derive 
the conditions for which the particular {\em vev} configurations used in this model
-- see Eqs. (\ref{vev1}), (\ref{vev2}) and (\ref{vev3}) and Table
\ref{tab1s} --
corresponds to the local minimum. For immediate reference the {\em vev}s are:
\begin{equation}
\langle \Phi^0 \rangle = \frac{v}{\sqrt{3}} \pmatrix{1 \cr 1 \cr 1} \;,\;
\langle \eta^0 \rangle = u  \;,
\langle \hat{\Delta}^{L0}_a \rangle = v_{La}\pmatrix{1 \cr 0 \cr 0} \;,\; 
\langle \hat{\Delta}^{L0}_b \rangle = v_{Lb}\pmatrix{1 \cr 1 \cr 1} \;,\; 
\langle \Delta_1^{L0} \rangle =  
\langle \Delta_2^{L0} \rangle =  
\langle \Delta_3^{L0} \rangle = {u}_L \;, 
\label{vev1a}
\end{equation}

\begin{equation} 
\langle \hat{\Delta}^{R0}_a \rangle = 
v_{Ra}\pmatrix{1 \cr 1 \cr 1} \;,\; 
\langle \hat{\Delta}^{R0}_b \rangle = 
v_{Rb}\pmatrix{1 \cr \omega \cr \omega^2} \;,\;
\langle \hat{\Delta}^{R0}_c \rangle = 
v_{Rc}\pmatrix{1 \cr \omega^2 \cr \omega} \;,\;
\label{vev2a}
\end{equation}

\begin{equation}
\langle \Delta_1^{R0} \rangle = {u}_{1R} \;,\; 
\langle \Delta_2^{R0} \rangle = {u}_{2R} \;,\; 
\langle \Delta_3^{R0} \rangle = {u}_{3R} \;.\; 
\label{vev3a}
\end{equation}

where the $SU(2)_L$ nature of the scalars has been suppressed.

Eq.
(\ref{vev1a}) shows that the $A4$ triplet fields --
$\hat{\Delta}^{L,R}$ and $\Phi$ -- have {\em vev} configurations that have
been verified to be the {\em global minima} in \cite{gmin1}. 
This result was for a single $A4$ triplet considered in isolation.
In the current scenario since many other
fields are involved, it is not 
straight-forward to directly adopt the conclusions of \cite{gmin1}. 

The conditions for which the {\em vev} configurations
shown in Eqs. (\ref{vev1}), (\ref{vev2}) and (\ref{vev3})
correspond to minimum are shown sector by sector.
\vskip 5pt
For minima 
of the scalar potential, the first derivatives of the scalar potential with respect to the {\em vev}s 
have to vanish and the second derivatives have to satisfy some conditions.
Since the scalar sector is very rich, the expressions look very complicated.
The conditions arising by setting the first derivatives to be zero have been
discussed for each of the $SU(2)_L$ sectors. As a sample, constraints coming from the
second derivatives have been shown only for the $SU(2)_L$ singlet sector.
Similar exercise can be carried out for the other $SU(2)_L$ sectors but are not presented here.

\subsubsection{$SU(2)_L$ singlet sector:}

The $SU(2)_L$ singlet {\em vev}s are much larger than those of the doublet
and triplet scalars. 
Thus it is safe to neglect the contributions to the minimization
equations from the inter-sector terms.

Let us remind ourselves that
$v_{Rp}$  $(p=a,b,c)$ are real and define:
\begin{eqnarray}
\tilde{v}_{Ra_1}\equiv v_{Ra},\ \ \tilde{v}_{Ra_2}\equiv v_{Ra},\ \ 
\tilde{v}_{Ra_3}\equiv v_{Ra};\nonumber\\
\tilde{v}_{Rb_1}\equiv v_{Rb},\ \ \tilde{v}_{Rb_2}\equiv v_{Rb}\omega, \ \
\tilde{v}_{Rb_3}\equiv v_{Ra}\omega^2;
\nonumber\\
\tilde{v}_{Rc_1}\equiv v_{Rc},\ \ \tilde{v}_{Rc_2}\equiv v_{Rc}\omega^2, \ \ \tilde{v}_{Rc_3}\equiv v_{Rc}\omega.
\end{eqnarray}
For ease of presentation, let us set the following masses and couplings equal:
\begin{eqnarray}
m^2_{\Delta_1^R}&=&m^2_{\Delta_2^R}=m^2_{\Delta_3^R}=m^2_{R1};\ \
m^2_{\hat{\Delta}_a^R}=m^2_{\hat{\Delta}_b^R}=m^2_{\hat{\Delta}_c^R}=m^2_{R2};\ \
m^2_{ab}=m^2_{ac}=m^2_{bc}=m^2_{R3};
\nonumber\\
\lambda^s_{1i}&=&\lambda^s_{1}\ \ \forall\ \ ( i=1,2,3); \ \ 
\lambda^s_{221}=\lambda^s_{231}=\lambda^s_{223}=\lambda^s_{2};\ \  
\lambda^s_{3a}=\lambda^s_{3b}=\lambda^s_{3c}=\lambda^s_{3} ;\ \
\widetilde{\lambda}^s_{3a}=\widetilde{\lambda}^s_{3b}=\widetilde{\lambda}^s_{3c}=\widetilde{\lambda}^s_{3};
\nonumber\\
\lambda^s_{3ab}&=&\lambda^s_{3ac}=\lambda^s_{3bc}=\hat{\lambda}^s_{3};\ \ 
\lambda^s_{4ip}=\lambda^s_{4}\ \ \forall\ \ (p=a,b,c)\  \ {\rm and} \ \ ( i=1,2,3); \ \ 
\nonumber\\
\lambda^s_{41ab}&=&\lambda^s_{41ac}=\lambda^s_{41bc}=\lambda^s_{42ab}=\lambda^s_{42ac}=\lambda^s_{42bc}
=\lambda^s_{43ab}= \lambda^s_{43ac}=\lambda^s_{43bc}=\widetilde{\lambda}^s_{4};
\nonumber\\
\lambda^s_{5a}&=&\lambda^s_{5b}=\lambda^s_{5c}=\lambda^s_{5}; \ \
\lambda^s_{5ab}=\lambda^s_{5ac}=\lambda^s_{5bc}=\widetilde{\lambda}^s_{5};\ \ 
\lambda^s_{6a}=\lambda^s_{6b}=\lambda^s_{6c}=\lambda^s_{6}; \ \
\lambda^s_{6ab}=\lambda^s_{6ac}=\lambda^s_{6bc}=\widetilde{\lambda}^s_{6}; \ \
\nonumber\\
\lambda^s_{7a}&=&\lambda^s_{7b}=\lambda^s_{7c}=\lambda^s_{7}; \ \
\lambda^s_{7ab}=\lambda^s_{7ac}=\lambda^s_{7bc}=\widetilde{\lambda}^s_{7};\ \ 
\lambda^s_{8ip}=\lambda^s_{8}\ \ \forall\ \ (p=a,b,c)\  \ {\rm and} \ \ ( i=1,2,3); \ \ 
\nonumber\\
\lambda^s_{81ab}&=&\lambda^s_{81ac}=\lambda^s_{81bc}=\lambda^s_{82ab}=\lambda^s_{82ac}=\lambda^s_{82bc}
=\lambda^s_{83ab}= \lambda^s_{83ac}=\lambda^s_{83bc}=\widetilde{\lambda}^s_{8};\ \ 
\nonumber\\
\lambda^s_{9ip}&=&\lambda^s_{9}\ \ \forall\ \ (p=a,b,c)\  \ {\rm and} \ \ ( i=1,2,3); \ \ 
\nonumber\\
\lambda^s_{91ab}&=&\lambda^s_{91ac}=\lambda^s_{91bc}=\lambda^s_{92ab}=\lambda^s_{92ac}=\lambda^s_{92bc}
=\lambda^s_{93ab}= \lambda^s_{93ac}=\lambda^s_{93bc}=\widetilde{\lambda}^s_{9}.
\label{se}
\end{eqnarray}
With the help of the singlet sector potential in Eq. (\ref{Vs}), the equalities in Eq. (\ref{se}) and the {\em
vev} in Eqs. (\ref{vev1}), (\ref{vev2}) and (\ref{vev3}) one can obtain:
\begin{eqnarray}
\frac{\partial V_{singlet}|_{min}}{\partial u_{1R}^*}=0
&\Rightarrow& m^2_{R1}u_{1R}+ \lambda^s_1 (u_{1R}^*u_{1R}^2)
+\lambda^s_2\left[(u_{2R}^*u_{2R})+(u_{3R}^*u_{3R})\right]\nonumber\\
&+& \frac{3\lambda_4^s}{2} u_{1R}\left[ v_{Ra}^2+ v_{Rb}^2+v_{Rc}^2\right] 
+3\lambda_5^s v_{Ra}^3\nonumber\\
&-& 3\widetilde{\lambda}_5^s v_{Ra}\left(v_{Rb}^2 + v_{Rc}^2\right)
+ 6 \lambda_8^s u_{1R}^* v_{Ra}^2 + 6 \widetilde{\lambda}_8^s u_{1R}^* v_{Rb}v_{Rc}\nonumber\\
&+& 3 \lambda_9^s \left(u_{3R}^* v_{Rb}^2+u_{2R}^* v_{Rc}^2  \right)
+3 \widetilde{\lambda}_9^s\left[ v_{Ra}\left(u_{2R}^* v_{Rb}+u_{3R}^* v_{Rc}  \right)\right]
=0
\;\;,
\end{eqnarray}

\begin{eqnarray}
\frac{\partial V_{singlet}|_{min}}{\partial \tilde{v}_{Ra_1}^* }&=& \frac{\partial V_{singlet}|_{min}}{\partial v_{Ra_1}^* }=0
\nonumber \\
& \Rightarrow & m_{R2}^2 v_{Ra}+ m_{R3}^2 \left(v_{Rb}+v_{Rc}\right)
+\frac{7}{2}\lambda^s_3 v_{Ra}^3 + \left(3 \hat{\lambda}_3 + \frac{\widetilde{\lambda}^s_3}{8}\right)
 v_{Ra}\left(v_{Rb}^2+v_{Rc}^2\right)\nonumber \\
&+&\left[\frac{\lambda^s_4}{2}v_{Ra}+ \widetilde{\lambda}^s_4(v_{Rb}+v_{Rc})\right]
\left(u_{1R}^*u_{1R}+u_{2R}^*u_{2R}+u_{3R}^*u_{3R}\right)
+\lambda^s_5 v_{Ra}^2 \left(u_{1R}^*+2u_{1R}\right)\nonumber \\
&-&\widetilde{\lambda}^s_5\left[(v_{Ra}v_{Rb}+v_{Ra}v_{Rc})(2 u_{1R}+u_{1R}^*)\right]
+\lambda^s_6 v_{Ra}^2 (u_{2R}^*-u_{2R})\nonumber \\
&+&\widetilde{\lambda}^s_6 \left[
u_{2R}\left(2v_{Rb}^2- v_{Rc}^2-v_{Ra}v_{Rb}+2v_{Rb}v_{Rc}\right)
-u_{2R}^*v_{Ra} (v_{Rb}+ v_{Rc})\right]+\lambda^s_7 v_{Ra}^2 (u_{3R}^*-u_{3R})\nonumber \\
&+&\widetilde{\lambda}^s_7 \left[
u_{3R}\left(2v_{Ra}v_{Rb}- v_{Ra}v_{Rc}-v_{Rb}^2+2v_{Rc}^2\right)
-u_{3R}^*v_{Ra} (v_{Rb}+ v_{Rc})\right]\nonumber \\
&+& \left(u_{1R}^2+u_{2R}^2+ u_{3R}^2\right)\left[2\lambda^s_8v_{Ra}+\widetilde{\lambda}^s_8(v_{Rb}+v_{Rc})\right]\nonumber \\
&+& \left(u_{1R}^*u_{2R}+ u_{1R}^*u_{3R}+u_{2R}^*u_{3R}\right)\left[2\lambda^s_9v_{Ra}+\widetilde{\lambda}^s_9(v_{Rb}+v_{Rc})\right]
 =0 \;\;,
\end{eqnarray}
Besides the first derivatives discussed above, second derivatives are also needed to established minimality. For example,
\begin{eqnarray}
\frac{\partial^2 V_{singlet}|_{min}}{\partial u_{1R}^{*2} }&>& 0
\nonumber \\
& \Rightarrow & \lambda^s_1 u_{1R}^2+6\lambda^s_8 v_{Ra}^2+ 6\widetilde{\lambda}^s_8v_{Rb}v_{Rc}>0
\label{bxx}
\end{eqnarray}
and
\begin{eqnarray}
\frac{\partial^2 V_{singlet}|_{min}}{\partial \tilde{v}_{Ra_1}^{*2} }&>&0 
\nonumber \\
& \Rightarrow & \lambda^s_3 v_{Ra}^2+ 4 \hat{\lambda}_3 \left(v_{Rb}^2+v_{Rc}^2\right)
+2\lambda^s_8 \left(u_{1R}^2+u_{2R}^2+ u_{3R}^2\right)\nonumber \\
&+&2\lambda^s_8 \left(u_{2R}u_{3R}+u_{1R}u_{2R}+u_{1R}u_{3R}\right)>0
\label{byy}
\end{eqnarray}
Further mixed derivatives such as: 
\begin{eqnarray}
\frac{\partial^2 V_{singlet}|_{min}}{\partial u_{1R}^*\tilde{v}_{Ra_1} }
& = & \frac{\lambda^s_4}{2}v_{Ra}u_{1R}+
\widetilde{\lambda}^s_4 \left(v_{Rb}+v_{Rc}\right)u_{1R}\nonumber \\
&+&\lambda^s_5v_{Ra}^2-\widetilde{\lambda}^s_5v_{Ra}\left(v_{Rb}+v_{Rc}\right)
\label{bcrs}
\end{eqnarray}
are also necessary to establish minimality in the most general case.
The results presented for the first and second derivatives are
calculated using the most general expression of the scalar potential in terms of the 
{\em vev}s and putting $(v_{Ra1}=v_{Ra2}=v_{Ra3}=v_{Ra})$, $(v_{Rb1}=v_{Rb},v_{Rb2}=\omega v_{Rb},v_{Rb3}=\omega^2 v_{Rb})$ and $(v_{Rc1}=v_{Rc}, v_{Rc2}=\omega^2v_{Rc}, v_{Rc3}=\omega v_{Rc})$ where $v_{Ra}$, $v_{Rb}$, $v_{Rc}$ are real.
Needless to mention that $\tilde{v}_{Rp_i}^*=\tilde{v}_{Rp_i}$ for ($p=a, i=1,2,3$) and
($p=b,c \ \ {\rm and}\ \ i=1$). Similar equations can be obtained by minimizing the 
potential wrt $u_{2R},\ \  u_{3R}$, $\tilde{v}_{Ra2}, \ \ \tilde{v}_{Ra3}$
and $\tilde{v}_{Rpi}$ for $(p=b,c)$ and $(i=1,2,3)$. For the sake of brevity those are not mentioned.
Similar strategy will be adopted for the $SU(2)_L$ doublet and $SU(2)_L$ triplet sector.
It is worth noting that this exercise for all the three sectors are for illustrative purpose only
and the minimization equations are achieved by setting the different couplings equal.
 
\subsubsection{$SU(2)_L$ doublet sector:}
For this sector contributions from both the
doublet sector itself -- Eq. (\ref{Vd}) -- together with the singlet-doublet inter-sector
are considered.
Let us define
$V_{\mathscr{D}}=V_{doublet}+V_{sd}$. Also let us call $\langle\Phi_i\rangle\equiv v_i$ where
$v_1=v_2=v_3=\frac{v}{\sqrt 3}$, $v$ being real.
\vskip 1pt
The following couplings are set to be equal:
\begin{eqnarray}
\lambda^{sd}_{1i}&=&\lambda^{sd}_{1},\ \  \lambda^{sd}_{2i}= \lambda^{sd}_{2}\ \ \forall\ \  (i=1,2,3);\nonumber\\
\lambda^{sd}_{3p}&=&\lambda^{sd}_{3},\ \  \lambda^{sd}_{4p}= \lambda^{sd}_{4},\ \ 
\lambda^{sd}_{5p}= \lambda^{sd}_{5},\ \ \lambda^{sd}_{6p}= \lambda^{sd}_{6}, 
\ \ \lambda^{sd}_{7p}= \lambda^{sd}_{7}
\ \ \forall \ \ (p=a,b,c);
\nonumber\\
\lambda^{sd}_{3ab}&=&\lambda^{sd}_{3ac}=\lambda^{sd}_{3bc}=\widetilde{\lambda}^{sd}_{3};\ \ 
\lambda^{sd}_{7ab}=\lambda^{sd}_{7ac}=\lambda^{sd}_{7bc}=\widetilde{\lambda}^{sd}_{7}.
\end{eqnarray}
For the {\em vev}s in Eqs. (\ref{vev1a}), (\ref{vev2a})
and (\ref{vev3a}) correspond to the minimum of the scalar
potential it is necessary to satisfy the following conditions:

\begin{equation}
\frac{\partial V_{\mathscr{D}}|_{min}}{\partial u^*}=0
\Rightarrow u \left[m_{\eta}^2+\lambda^d_1 u^*u +\lambda^d_3 v^2
+\lambda^{sd}_{1} \sum_{i=1}^3\left(u_{iR}^*u_{iR}\right)
+\frac{3}{2}\lambda^{sd}_{3} \sum_{p=a}^c{v}_{Rp}^2 \right]
=0.
\label{DMIN2}
\end{equation}
and
\begin{eqnarray}
\frac{\partial V_{\mathscr{D}}|_{min}}{\partial v_1^* }&=&0
\nonumber\\
&\Rightarrow&\frac{v}{\sqrt{3}}\left[ m_{\Phi}^2+
2\lambda_2^d\frac{v^2}{3}+\frac{\lambda_3^d}{2}(u^*u)
+\frac{\lambda^{sd}_{2}}{2} \sum_{i=1}^3\left(u_{iR}^*u_{iR}\right)
\right.
\nonumber\\
&+& \left. 
\frac{\lambda^{sd}_{4}}{2}\left(2v_{Ra}-v_{Rb}-v_{Rc}\right)\left(u_{1R}+u_{1R}^*\right)
\right.
\nonumber\\
&+& \left.
\frac{\lambda^{sd}_{7}}{2}
\left[\left(\sum_{p=a}^c 3v_{Rp}^2\right)+\frac{1}{2}\left(2v_{Ra}^2-v_{Rb}^2-v_{Rc}^2\right)
\right]
\right.
\nonumber\\
&+& \left.
 \frac{\widetilde{\lambda}^{sd}_{7}}{2}\left[6\left(v_{Ra}v_{Rb}+v_{Ra}v_{Rc}+v_{Rb}v_{Rc}\right)
+\frac{1}{2}\left(v_{Ra}v_{Rb}+v_{Ra}v_{Rc}-2v_{Rb}v_{Rc}\right)\right]
\right]=0.
\nonumber\\
\label{DMIN3}
\end{eqnarray}

In order
to satisfy Eqs. (\ref{DMIN2}) and (\ref{DMIN3}) some degree of fine-tuning
is necessary that involve
both $SU(2)_L$ doublet and singlet {\em vev} of varying magnitudes.
Similar equations can be obtained by minimizing the potential wrt $v_2^*$ and $v_3^*$.

\subsubsection{$SU(2)_L$ triplet sector:}

In analogy to the doublet sector, let us define 
$V_{\mathscr{T}}=V_{triplet}+V_{ts}$ using Eqs. (\ref{Vt}) and (\ref{Vts}).
Let us also recall, $v_{La1}=v_{La}$, $v_{La2}=v_{La3}=0$
and $v_{Lb1}=v_{Lb2}=v_{Lb3}=v_{Lb}$. 

This sector has several couplings involved. For simplicity
of presentation, let us implement the following choices:
\begin{eqnarray}
m_{\Delta^L_1}&=&m_{\Delta^L_2}=m_{\Delta^L_3}=m_{t1}; \ \ 
m_{\hat\Delta^L_a}=m_{\hat\Delta^L_b} = m_{t2}; \ \  m_{ab}=m_{t3};\ \  
\lambda^t_{1i}=\lambda^t_{1}, \ \ \forall \ \ (i=1,2,3) ;
\ \ \nonumber\\
\lambda^t_{221}&=&\lambda^t_{232}=\lambda^t_{231}=\lambda^t_{2} \;\;; \;\;
\lambda^t_{3a}=\lambda^t_{3b}=\lambda^t_{3}; \ \ \lambda^t_{3ab}=\hat{\lambda}^t_{3}
; \ \ \widetilde{\lambda}^t_{3ab}=\widetilde{\lambda}^t_{3}; \ \
\nonumber\\
\lambda_{4jn}&=&\lambda_{4jnl}=\lambda_{4}
; \ \
\lambda_{8jn}=\lambda_{8jnl}=\lambda_{8}
; \ \
\lambda_{9jn}=\lambda_{9jnl}=\lambda_{9}, \ \ \forall \ \ (j=1,2,3), \ \ (n,l=a,b) \ \ {\rm and}\ \ n\ne l ;
\ \ \nonumber\\
\lambda^t_{5a}&=&\lambda^t_{5b}=\lambda^t_{5ab}=\lambda^t_{5};\ \ 
\lambda^t_{6a}=\lambda^t_{6b}=\lambda^t_{6ab}=\lambda^t_{6};\ \ 
\lambda^t_{7a}=\lambda^t_{7b}=\lambda^t_{7ab}=\lambda^t_{7};\ \ \nonumber\\
\lambda^{ts}_{1ij}&=&\lambda^{ts}_1 , \ \ \forall \ \ (i,j=1,2,3)\ \  {\rm and}\ \ i\ne j;
\ \ 
\lambda^{ts}_{2jn}=\lambda^{ts}_{2jnl}, \ \ \forall \ \ (j=1,2,3), \ \ (n,l=a,b) \ \ {\rm and}\ \ n\ne l ;
\ \ \nonumber\\
\lambda^{ts}_{3pi}&=&\lambda^{ts}_{3pqi}, \ \ 
\forall \ \ (i=1,2,3), \ \ (p,q=a,b,c) \ \ {\rm and}\ \ p\ne q ;
\ \ \nonumber\\
\lambda^{ts}_{4jnnpp}&=&\lambda^{ts}_{4jnlpp}=\lambda^{ts}_{4jnnpq}=\lambda^{ts}_{4jnlpq}
=\lambda^{ts}_4, \, 
\forall \,(j=1,2,3), \,(p,q=a,b,c), \, (n,l=a,b)\, {\rm and}\, p\ne q,\, n\ne l ;
\ \ \nonumber\\
\lambda^{ts}_{5jppn}&=&\lambda^{ts}_{5jpqn}=\lambda^{ts}_5, \ \ 
\forall \ \ (j=1,2,3), \ \ (p=a,b,c), \ \ (n=a,b);\nonumber\\
\lambda^{ts}_{6jnnp}&=&\lambda^{ts}_{6jnlp}=\lambda^{ts}_6, \ \ 
\forall\ \ (j=1,2,3), \ \ (p=a,b,c),\ \ (n,l=a,b) \ \ {\rm and}\ \ n\ne l. 
\nonumber\\
\label{TMIN1}
\end{eqnarray}
In order to minimize $V_{\mathscr{T}}$ such that one can arrive to
{\em vev}s furnished in Eqs.  (\ref{vev1a}), (\ref{vev2a})
and (\ref{vev3a}) the following conditions are to be ensured:
\begin{eqnarray}
\frac{\partial V_{\mathscr{T}}|_{min}}{\partial u_{L1}^*}&=&0\nonumber\\
&\Rightarrow& u_L\left[m_{t1}^2 + (u_L^*u_L)\left(\lambda_1^t+\lambda_2^t\right)
+\frac{\lambda_4^t}{2}\left(v_{La}^2+3v_{Lb}^2+2v_{La}v_{Lb}\right)
\right]
+\lambda_5^t v_{Lb}^2\left(3v_{Lb}+v_{La}\right)
\nonumber\\
&+& 2\lambda_8^t u_L^* \left(v_{La}^2+3v_{Lb}^2+v_{La}v_{Lb}\right)
+2\lambda_9^t v_{La} \left(v_{La}+v_{Lb}\right)
+\lambda_1^{ts} u_L\sum_{i=1}^3 \left(u_{iR}^*u_{iR}\right)\nonumber\\
&+& \frac{3}{2}\lambda_3^{ts} u_L\left[v_{Ra}^2+v_{Rb}^2+v_{Rc}^2+v_{Rb}v_{Rc}\right]
\nonumber\\
&+&\lambda_5^{ts} \left(v_{La}+3v_{Lb}\right)\left[2v_{Ra}^2-v_{Rb}^2-v_{Rc}^2-2v_{Ra}^2(
v_{Ra}+v_{Rb}^2) + 4v_{Rb}v_{Rc}\right]\nonumber\\
&+& \left[\left(\lambda_7^ts u_{1R}+\lambda_{12}^ts u_{1R}^*\right)\left[
v_{La}\left(v_{Ra}+v_{Rb}+v_{Rc}\right)
+3v_{Ra}v_{Lb}\right]
\right]
\nonumber\\
&+& \left[\left(\lambda_8^ts u_{3R}+\lambda_{10}^ts u_{3R}^*\right)\left[
v_{La}\left(v_{Ra}+v_{Rb}+v_{Rc}\right)
+3v_{Rc}v_{Lb}\right]
\right]
\nonumber\\
&+& \left[\left(\lambda_9^ts u_{2R}+\lambda_{11}^ts u_{2R}^*\right)\left[
v_{La}\left(v_{Ra}+v_{Rb}+v_{Rc}\right)
+3v_{Rb}v_{Lb}\right]
\right]
=0.
\label{DMIN22}
\end{eqnarray}

Also one gets;
\begin{eqnarray}
\frac{\partial V_{\mathscr{T}}|_{min}}{\partial v_{La_1}^*}&=&0\nonumber\\
&\Rightarrow&  v_{La}\left[m_{t1}^2+m_{t2}^2+2\lambda_3^{t}v_{La}^2
+4\hat{\lambda}^t_3v_{Lb}^2+\frac{3}{2}\widetilde{\lambda}_3^{t}v_{Lb}^2
\right]
+\frac{3}{2}\lambda_4^{t} \left(u_L^*u_L\right)\left(v_{La}+v_{Lb}\right)
\nonumber\\ 
&+&\left( 2\lambda_5^{t}-\lambda_6^t -\lambda_7^t \right) v_{Lb}^2u_L
+\lambda_8^t u_L^2 \left(2v_{La}+3v_{Lb}\right)
+\lambda_9^t u_L^2 \left(2v_{La}+v_{Lb}\right)\nonumber\\ 
&+&\frac{\lambda_2^{ts}}{2}\left[\left(v_{La}+v_{Lb}\right)\sum_{i=1}^3u_{iR}^*u_{iR}\right]
+\frac{\lambda_4^{ts}}{2}\left[3\left(v_{La}+v_{Lb}\right)\sum_{p=a}^cv_{Rp}^2+
\frac{1}{2}v_{Lb}\left(2v_{Ra}^2-v_{Rb}^2-v_{Rc}^2\right)\right]\nonumber\\ 
&+&\widetilde{\lambda}_4^{ts}\left[3\left(v_{La}+v_{Lb}\right)\left(v_{Ra}v_{Rb}+v_{Rb}v_{Rc}\right)
+\frac{1}{2}v_{Lb}\left(v_{Ra}v_{Rb}+v_{Ra}v_{Rc}+v_{Rb}v_{Rc}\right)\right]
\nonumber\\ 
&+&\lambda_5^{ts}\left[3u_L\left(2v_{Ra}^2-v_{Rb}^2-v_{Rc}^2
-2v_{Ra}v_{Rb}-2v_{Ra}v_{Rc}+4v_{Rb}v_{Rc}\right)
\right]\nonumber\\ 
&+& \left[u_L \left(v_{Ra}+v_{Rb}+v_{Rc}\right)\right]
\left[ \left[\left(u_{1R}^*+u_{2R}^*+u_{3R}^*\right)\left(\lambda_7^{ts}+\lambda_8^{ts}+\lambda_9^{ts}\right)\right] 
\right.
\nonumber\\
&+& \left.
\left[\left(u_{1R}+u_{2R}+u_{3R}\right)\left(\lambda_{10}^{ts}+\lambda_{11}^{ts}+\lambda_{12}^{ts}\right)\right]\right]
=0.
\nonumber\\
\label{DMIN32}
\end{eqnarray}
It is worth noticing that certain fine-tuning is essential to satisfy
Eqs. (\ref{DMIN22}) - (\ref{DMIN32}). Also similar equations can be obtained by
minimizing the potential wrt $u_{Lj}^*$ where $(j=2,3)$, $v_{Lni}^*$ where for $n=b$ one has $(i=1,2,3)$  
and for $n=a$ we have $(i=2,3)$. Those are not mentioned here. This exercise is performed to illustrate 
the scenario in a simplified limit achieved by setting several masses and couplings to be equal.



\setcounter{equation}{0}  
\section{Appendix: Flavour basis form of the mass matrices}\label{MassA4}
Mass matrices expressed in the Lagrangian basis 
in Eqs. (\ref{mmatrix1}) and (\ref {mmatrix2}) can be 
transformed to simpler forms in the flavour basis as in 
Eqs. (\ref{mflav1}) and (\ref{mflav1a})
with the help of a unitary transformation written in Eq. (\ref{chngbasis}).
Certain straight-forward algebraic calculations related to this
derivation of the forms the mass matrices in the flavour basis 
is furnished in this Appendix.
\vskip 1pt
The Lagrangian in Eq. (\ref{e1}) produces
the following mass matrix for the charged
leptons and the left-handed Majorana neutrinos:
\begin{equation}
M_{e\mu\tau}=\frac{v}{\sqrt{3}}
\pmatrix{y_1 & y_2 & y_3\cr
y_1 & \omega y_2 & \omega^2 y_3\cr
y_1 & \omega^2 y_2 & \omega y_3} \;,\;M_{\nu L}=
\pmatrix{(Y^L_1+ 2 Y^L_2)u_L & \frac{1}{2}\hat{Y}^L_b v_{Lb}& \frac{1}{2}\hat{Y}^L_b v_{Lb} \cr
\frac{1}{2}\hat{Y}^L_b v_{Lb} & (Y^L_1 - Y^L_2)u_L & \frac{1}{2}(\hat{Y}^L_a v_{La} +\hat{Y}^L_b v_{Lb} )\cr
\frac{1}{2}\hat{Y}^L_b v_{Lb} & \frac{1}{2}(\hat{Y}^L_a v_{La} +\hat{Y}^L_b v_{Lb} ) & (Y^L_1 - Y^L_2)u_L},
\label{mmatrix1app}
\end{equation}
where, the Yukawa coupling $Y^L_2$ is chosen to be equal to $Y^L_3$. Also, $y_1 v = m_e, ~ y_2 v = m_\mu , ~ y_3 v = m_\tau$ is satisfied.
The dominant Type-II see-saw component of the neutrino mass matrix, $M_{\nu L}$,
gives rise to the atmospheric splitting and maximal atmospheric mixing but is devoid of 
solar splitting and is therefore characterized by two masses $m_1^{(0)}$
and $m_3^{(0)}$. It is useful to define $m^\pm\equiv m_3^{(0)}\pm m_1^{(0)}$.
Thus $m^-$ is positive (negative) for normal (inverted) ordering.
Certain identifications of the 
{\em vev} and Yukawa products are essential viz. $3(Y^L_1+ 2 Y^L_2)u_L=(m_3^{(0)}+m^+)$, $6(Y^L_1 - Y^L_2)u_L=\hat{Y}^L_a v_{La}=m^+$ and
$3\hat{Y}^L_b v_{Lb}=-2m^-$ to generate the 
desired structures of the mass matrices as presented in Eq. (\ref{mflav1}).  
The neutrino Dirac mass matrix and the 
right-handed Majorana neutrino mass matrix in the Lagrangian basis are:
\begin{equation}
M_D = f u ~\mathbb{I}\;\;,\;\; 
M_{\nu R}= m_R\pmatrix{\chi_1& 
\chi_6 &\chi_5\cr 
\chi_6& \chi_2 & \chi_4\cr 
\chi_5 &\chi_4 & \chi_3}\;\;,
\label{mmatrix2app}
\end{equation}
where, 
\begin{eqnarray}
m_R\chi_1&\equiv&(Y^R_1 u_{1R}+Y^R_2 u_{2R}+Y^R_3 u_{3R}) \nonumber\\
m_R\chi_2&\equiv&(Y^R_1 u_{1R}+\omega Y^R_2 u_{2R}+\omega^2 Y^R_3 u_{3R})\nonumber\\
m_R\chi_3&\equiv&(Y^R_1 u_{1R}+\omega^2 Y^R_2 u_{2R}+\omega Y^R_3 u_{3R})\nonumber\\
m_R\chi_4&\equiv&\frac{1}{2}(\hat Y^R_a v_{Ra}+\hat Y^R_b v_{Rb} +\hat Y^R_c v_{Rc}) \nonumber\\
m_R\chi_5&\equiv&\frac{1}{2}(\hat Y^R_a v_{Ra}+\omega\hat Y^R_b v_{Rb} + \omega^2\hat Y^R_c v_{Rc}) \nonumber\\
m_R\chi_6&\equiv&\frac{1}{2}(\hat Y^R_a v_{Ra}+\omega^2\hat Y^R_b v_{Rb} + \omega\hat Y^R_c v_{Rc}).
\label{mmatrix2chidef}
\end{eqnarray}
Here $m_R$ is the right-handed Majorana neutrino mass scale
and $\chi_i$ are dimensionless ${\cal O}(1)$ quantities. 
In order to achieve the right-handed Majorana neutrino mass matrix of the form expressed in 
Eq. (\ref{mflav1a}), the {\em vev} and Yukawa 
couplings products have to obey:
\begin{eqnarray}
Y^R_1 u_{1R}=m_R (r_{11}+2r_{23}),\ \ Y^R_2 u_{2R}= m_R(r_{22}+2r_{13}), 
\ \ Y^R_3 u_{3R}= m_R(r_{33}+2r_{12})\nonumber\\
\hat Y^R_a v_{Ra}=2m_R(r_{11}-r_{23}), \ \ \hat Y^R_b v_{Rb}=2m_R(r_{22}-r_{13}) \ \ 
{\rm and} \ \ \hat Y^R_c v_{Rc}=2m_R(r_{33}-r_{12}).
\label{mmatrix2vevyukawa}
\end{eqnarray}
The $r_{ij}$ in Eq. (\ref{mmatrix2vevyukawa}) are given by :
\begin{eqnarray}
r_{11}&\equiv&\sqrt{2}b \sin 2\theta_{12}^0+a \sin^2\theta_{12}^0,\ \ \nonumber\\
r_{22}&\equiv&-\sqrt{2}b \sin\theta_{12}^0-\frac{b}{2} \sin 2\theta_{12}^0
-a\cos\theta_{12}^0+\frac{a}{2} \cos^2\theta_{12}^0+\frac{a}{2}, \ \ \nonumber\\
r_{33}&\equiv&-\frac{b}{\sqrt{2}} \sin 2\theta_{12}^0- \sqrt{2}b\sin \theta_{12}^0
+a\cos\theta_{12}^0+\frac{a}{2} \cos^2\theta_{12}^0+\frac{a}{2},
\nonumber\\
r_{12}&\equiv&b \cos 2\theta_{12}^0+ \frac{a}{2\sqrt{2}} \sin^2\theta_{12}^0
+b \cos\theta_{12}^0-\frac{a}{\sqrt{2}} \sin\theta_{12}^0,\ \ \nonumber\\
r_{13}&\equiv& -b \cos 2\theta_{12}^0-\frac{a}{2\sqrt{2}} \sin^2\theta_{12}^0
+b \cos\theta_{12}^0-\frac{a}{\sqrt{2}} \sin\theta_{12}^0,\ \ \nonumber\\
r_{23}&\equiv& \frac{b}{2} \sin 2\theta_{12}^0 -\frac{a}{2} \cos^2\theta_{12}^0+\frac{a}{2}
\ \ .
\label{rij}
\end{eqnarray}
 where $a$ and $b$ are dimensionless quantities of ${\cal O}(1)$.
The charged lepton mass matrix is not diagonal in the Lagrangian basis.
In order to go to a basis in which the charged lepton mass matrix $M_{e\mu\tau}$ is diagonal
a unitary transformation 
$U_L$ is applied on the 
left-handed lepton doublets. The transformation $V_R$ is applied on the right-handed neutrino singlets of $SU(2)_L$ such that the Dirac neutrino mass matrix remains proportional to identity in this transformed basis as well. This basis in which the charged lepton mass matrix is diagonal and the entire lepton mixing
is dictated by the neutrino sector is called the {\em flavour basis}.
The right-handed charged leptons were kept unchanged. The transformation matrices 
are given by:
\begin{equation}
U_L = {1 \over{\sqrt 3}}
\pmatrix{1 & 1 & 1 \cr
1 & \omega^2 &  \omega\cr
1 &  \omega&  \omega^2} = V_R \;.
\label{chngbasisapp}
\end{equation}
The mass matrices in the flavour basis are:
\begin{equation}
M_{e\mu\tau}^{flavour} = \pmatrix{m_e & 0 & 0 \cr 0 & m_\mu & 0 \cr
0 & 0 & m_\tau}\;\;,\;\;  
M_{\nu L}^{flavour} = {1 \over 2}
\pmatrix{2m^{(0)}_1 & 0 & 0 \cr 0 & m^+ & m^- \cr 0 & m^- & m^+} \;\;,
\label{mflav1app}
\end{equation}
\begin{equation}
M_D = f u ~\mathbb{I}\;\;,\;\;  
M_{\nu R}^{flavour} = {m_R \over 4ab}
\pmatrix{r_{11} & r_{12} & r_{13} \cr r_{12}& r_{22} & r_{23} \cr r_{13} & r_{23} & r_{33}} 
\;\;.
\label{mflav1aapp}
\end{equation}
One can identify $fu=m_D$ where $m_D$ is the scale of the  Dirac masses of the neutrinos. 
Type-I see-saw mechanism contribution is given by the matrices in Eq. (\ref{mflav1aapp}).





\end{document}